\documentclass[preprint]{sig-alternate-05-2015}
\toappear{[To appear in SIGMOD 2017.]}
\usepackage{times}
\usepackage{amssymb}
\usepackage{graphicx}
\usepackage{balance}
\usepackage{color}
\usepackage{graphicx}
\usepackage{caption}
\usepackage{amsmath}
\usepackage{listings}
\usepackage{epstopdf}
\usepackage{enumitem}
\usepackage{hhline}
\usepackage{multirow}
\usepackage[dvipsnames]{xcolor}
\usepackage{soul}
\usepackage{subfig}
\usepackage{hyperref}

\soulregister\cite7
\soulregister\ref7
\soulregister\pageref7

\usepackage[ruled,vlined]{algorithm2e}
\definecolor{dkgreen}{rgb}{0,0.6,0}
\definecolor{gray}{rgb}{0.5,0.5,0.5}
\definecolor{mauve}{rgb}{0.58,0,0.82}
\usepackage[capitalize,noabbrev]{cleveref}

\begin{document}
\sloppy



\clubpenalty=10000 
\widowpenalty = 10000


\title{Fast Failure Recovery for \\Main-Memory DBMSs on Multicores}

\newcommand{\system}{\textsc{Pacman}\xspace}
\newcommand{\systemcaption}{PACMAN\xspace}
\newcommand{\database}{Peloton\xspace}
\newcommand{\expfigsize}{0.7}
\definecolor{color1}{rgb}{0.8,.95,1}
\definecolor{color2}{rgb}{1,.8,0.6}

\definecolor{todo-color}{rgb}{1,0,0}
\newcommand{\todo}[1]{\textnormal{\color{todo-color}{\textbf{#1}}}\unskip}

\newcommand{\ccy}[1]{\textcolor{red}{{\bf #1}}}
\newcommand{\Removed}[1]{}

\newcommand{\eat}[1]{}

\numberofauthors{1}
\author{
\alignauthor
Yingjun Wu, Wentian Guo, Chee-Yong Chan, Kian-Lee Tan\\
\affaddr{School of Computing, National University of Singapore}\\
\email{\{yingjun, wentian, chancy, tankl\}@comp.nus.edu.sg}
}

\maketitle

\begin{abstract}

Main-memory database management systems (DBMS) can 
achieve excellent performance when processing 
massive volume of on-line transactions 
on modern multi-core machines. 
But existing durability schemes, namely,
tuple-level and transaction-level
logging-and-recovery mechanisms, 
either degrade the performance of transaction processing 
or slow down the process of failure recovery. 
In this paper, we show that, by exploiting application semantics, 
it is possible to achieve speedy failure recovery without 
introducing any costly logging overhead to the 
execution of concurrent transactions. 
We propose \system, a parallel database recovery mechanism 
that is specifically designed for lightweight, coarse-grained 
transaction-level logging. 
\system leverages a combination of static and dynamic analyses to
parallelize the log recovery: 
at compile time, \system decomposes stored procedures by 
carefully analyzing dependencies within and across programs; 
at recovery time, \system exploits the availability of the runtime
parameter values to attain an execution schedule 
with a high degree of parallelism.
As such, recovery performance is remarkably increased. 
We evaluated \system in a fully-fledged main-memory DBMS 
running on a 40-core machine. 
Compared to several state-of-the-art database recovery mechanisms, 
\system can significantly reduce recovery time without 
compromising the efficiency of transaction processing.

\end{abstract}

\section{Introduction}
\label{introduction}

The on-going evolution of modern computer architectures has led to the 
rapid development of main-memory DBMSs. 
By resolving potential performance bottlenecks such as disk accesses and 
centralized contention points, modern main-memory DBMSs can power 
OLTP applications at very high throughput of millions of 
transactions per second on a multi-core machine~\cite{johnson2009shore,kallman2008h,kemper2011hyper,tu2013speedy}.

However, system robustness can be the Achilles' heel of such DBMSs. 
To preserve durability, a DBMS continuously persists
transaction logs during execution to ensure that the database 
can be restored to a consistent state after a failure, 
with all the committed transactions reflected correctly.

Existing approaches for DBMS logging can be broadly 
classified into two categories, each characterized by different 
granularities and performance emphasis.
Originally designed for disk-based DBMSs,
\textit{tuple-level logging} schemes, 
which include \textit{physical logging} (a.k.a. data logging)
and \textit{logical logging} (a.k.a. operation logging)\footnote{
In this paper, we follow the definitions presented in \cite{gray1992transaction}.},
propagate every tuple-level modification 
issued from a transaction to the secondary storage 
prior to the transaction's final commitment~\cite{mohan1992aries}. 
Such a heavyweight, fine-grained approach can generate 
tens-of-gigabyte of logging data per minute, 
causing over 40\% performance degradation for transaction execution 
in a fast main-memory DBMSs~\cite{malviya2014rethinking,zheng2014fast}. 
However, from the perspective of database recovery, 
tuple-level log recovery can be easily performed in parallel, 
and the recovery time can be further reduced by applying the 
last-writer-wins rule (a.k.a. Thomas write rule~\cite{zheng2014fast}). 
As an alternative to tuple-level logging, 
\textit{transaction-level logging}, 
or \textit{command logging}~\cite{malviya2014rethinking},
is initially invented for main-memory DBMSs that 
leverage deterministic
execution model for processing transactions~\cite{kallman2008h,stonebraker2007end,thomson2012calvin}. 
In contrast to common practice, most transactions in this type of DBMSs are issued from predefined
\textit{stored procedures}. In this scenario,
transaction-level logging can simply dump transaction logic, 
including a stored procedure identifier and the corresponding query parameters,
into secondary storage.
This coarse-grained strategy incurs very low overhead to 
in-memory transaction processing.
However, it also significantly slows down the recovery process,
as transaction-level log recovery is 
widely believed to be hard to 
parallelize~\cite{malviya2014rethinking,zheng2014fast}. 
To achieve high performance in both transaction processing and 
failure recovery, recent efforts have largely focused
on exploiting new hardware (e.g., non-volatile memory) 
to minimize the runtime overhead caused by 
tuple-level logging~\cite{johnson2010aether,ongaro2011fast,wang2014scalable,zheng2014fast}.

In this paper, we present \mbox{\system}, a parallel failure recovery mechanism 
that is specifically designed for 
lightweight, coarse-grained transaction-level logging in the context of
main-memory multi-core DBMSs.
The design of \system is inspired by two observations.
First, DBMSs utilizing transaction-level logging 
issue transactions from stored procedures.
This allows \system to analyze the stored procedures to 
understand the application semantics.
Second, DBMSs recover lost database states by re-executing transactions
in their original commitment order,
and this order is determined before system crash.
This allows \system to parallelize transaction-level log recovery
by carefully leveraging the dependencies within and across transactions.

\system models the transaction-level log recovery
as a \textit{pipeline of data-flow processing}.
This is accomplished
by incorporating a combination of static and dynamic analyses. 
At compile time, \system conservatively decomposes a collection of 
stored procedures into multiple conflict-free units, which are organized into 
a dependency graph that captures potential 
\textit{happen-before} relations. 
This prior knowledge enables fast transaction-level log recovery 
with a high degree of parallelism,
and this is achieved by generating an execution schedule 
through exploiting the availability of the runtime parameter values
of the lost transactions.

Unlike many state-of-the-art database logging-and-recovery schemes~\cite{johnson2010aether,ongaro2011fast,wang2014scalable,zheng2014fast}, 
\system does not make any assumption on the performance of the underlying hardware. 
It is also orthogonal to data layouts
(e.g., single-version or multi-version, row-based or column-based) and 
concurrency control schemes (e.g., two-phase locking or timestamp ordering), 
and can be applied to many main-memory DBMSs, 
such as Silo~\cite{tu2013speedy} and Hyper~\cite{kemper2011hyper}. 
\system's analysis approach also departs far from the existing, purely static, 
program partitioning and transformation techniques~\cite{cheung2012automatic,pandis2010data,ramachandra2012program,shasha1995transaction}, 
in that \system yields a program decomposition that is especially tailored for 
the execution of pre-ordered transaction sequences, 
and a higher degree of parallelism is attained by incorporating runtime information 
during failure recovery.

In contrast to the existing transaction-level 
log recovery mechanism~\cite{malviya2014rethinking} 
that relies on partitioned data storage for parallelization
(i.e., two transaction-level logs from different transactions accessing different 
data shards could be replayed in parallel),
\system is the first parallel recovery mechanism for 
transaction-level logging scheme that goes beyond partitioned-data parallelism.
Specifically, \system innovates with a combination of static and dynamic analyses 
that enable multiple recovery operations to be parallelized even when accessing
the same data shard.

We implemented \system as well as several state-of-the-art 
recovery schemes in \database~\cite{peloton}, 
a fully fledged main-memory DBMS 
optimized for high-performance multi-core transaction processing. 
Through a comprehensive experimental study,
we spotted several performance bottlenecks of 
existing logging-and-recovery schemes for
main-memory DBMSs,
and confirmed that \system can significantly reduce recovery time without 
bringing any costly overhead to transaction processing.

We organize the paper as follows: 
\cref{preliminary} reviews durability techniques for main-memory DBMSs.
\cref{overview} provides an overview of \system.
\cref{recovery} demonstrates how \system achieves fast failure recovery 
with a combination of static and dynamic analyses.
\cref{discussion} discusses the potential limitations of \system.
We report extensive experiment results in
\cref{evaluations}. 
\cref{relatedwork} reviews related works and \cref{conclusion} concludes.

\section{DBMS durability}
\label{preliminary}

A main-memory DBMS employs \textit{logging} and \textit{checkpointing} mechanisms 
during transaction execution to guarantee the durability property.

\subsection{Logging}
A main-memory DBMS continuously records transaction changes into 
secondary storage so that the effects of committed transactions 
can persist even in the midst of system crash. 
Based on the granularity, existing logging mechanisms for main-memory DBMSs 
can be broadly classified into two categories: 
\textit{tuple-level logging} and \textit{transaction-level logging}. 

Initially designed for disk-based DBMSs,
tuple-level logging keeps track of the images of modified tuples
and persists them into secondary storage before the transaction results
are returned to the clients.
According to the types of log contents, 
tuple-level logging schemes can be further classified into two sub-categories:
(1) \textit{physical logging}, which records the physical
addresses and the corresponding tuple values modified by a transaction;
and (2) \textit{logical logging}, which persists the write actions and the parameter
values
of each modification issued by a transaction.
Although logical logging usually generates smaller log records compared to
physical logging,
its assumption of \textit{action consistency}~\cite{gray1992transaction},
which requires each logical operation to be either completely done or
completely undone,
renders it unrealistic for disk-based DBMSs. 
Hence, many conventional disk-based DBMSs including MySQL~\cite{mysql} and Oracle~\cite{oracle}
adopt a combination of physical logging and logical logging,
or called \textit{physiological logging}, to minimize log size while addressing
action inconsistency problem.
While disk-based DBMS leverages \textit{write-ahead logging} 
to persist logs before the modification is applied to the database state,
main-memory DBMSs can delay the persistence of these log records 
until the commit phase of a transaction~\cite{diaconu2013hekaton,zheng2014fast}.
This is because such kind of DBMSs maintain all the states in memory, 
and dirty data is never dumped into secondary storage.
This observation makes it possible to
record only after images of all the modified tuples for a main-memory DBMS,
and logical logging can be achieved, as the action inconsistency problem
in disk-based DBMSs never occurs in the main-memory counterparts.

Transaction-level logging, or \textit{command logging}, 
is a new technique that is initially designed
for deterministic main-memory DBMSs~\cite{malviya2014rethinking}.
As this type of DBMSs require the applications to issue transactions as stored procedures,
the logging component in such a DBMS therefore only needs to 
record coarse-grained transaction logic, 
including the stored procedure identifier and the corresponding parameter values, 
into secondary storage; updates of any 
aborted transactions are discarded without being persisted.
A well-known limitation of transaction-level logging is that the recovery
time can be much higher compared to traditional tuple-level logging schemes,
and existing solutions resort to replication techniques to mask single-node
failures. 
The effectiveness of this mechanism, however, is heavily dependent 
on the networking speed, which in many circumstances (e.g., geo-replicated) 
is unpredictable~\cite{Corbett2012Spanner}.

A major optimization for DBMS logging is 
called \textit{group commit}~\cite{dewitt1984implementation,gawlick1985varieties},
which groups multiple log records into a single large I/O so as to 
minimize the logging overhead brought by frequent disk accesses.
This optimization is widely adopted in both disk-based and main-memory DBMSs.

\subsection{Checkpointing}
A main-memory DBMS periodically persists its table space 
into secondary storage to bound the maximum recovery time. 
As logging schemes in main-memory DBMSs do not record before images of modified tuples,
these DBMSs must perform transactionally-consistent checkpointing 
(rather than fuzzy checkpointing~\cite{li1993post})
to guarantee the recovery correctness.
Retrieving a consistent snapshot in a multi-version DBMS is straightforward,
as the checkpointing threads in this kind of DBMSs can 
access an older version of a tuple
in parallel with any active transaction, even if the transaction is modifying the same tuple.
However, for a single-version DBMS, checkpointing must be explicitly 
made asynchronous without
blocking on-going transaction execution~\cite{kallman2008h,kemper2011hyper,zheng2014fast}.

The checkpointing scheme in a DBMS must be compatible with 
the adopted logging mechanism.
While physical logging requires the checkpointing threads to persist 
both the content and the location of each tuple in the database,
logical logging and command logging only require 
recording the tuple contents during checkpointing.

\subsection{Failure Recovery}
A main-memory DBMS masks outages using persistent checkpoints and recovery logs.
Once a system failure occurs, the DBMS recovers the most recent transactionally-consistent 
checkpoint from the secondary storage. 
To recover the checkpoints persisted for physical logging, the DBMS only
needs to restore the table space, and the database indexes can be reconstructed
lazily at the end of the subsequent log recovery phase. 
However, recovering the checkpoints persisted
for logical logging and command logging requires the DBMS to reconstruct the database indexes simultaneously with the table space restoration.
After checkpoint recovery completes, 
the DBMS subsequently reloads and replays the durable log sequences according to the transaction commitment order, in which manner the DBMS can reinstall
the lost updates of committed transactions correctly.

\subsection{Performance Trade-Offs}
Based on the existing logging-and-recovery mechanisms, it is difficult to 
achieve high performance in both transaction processing and failure recovery 
in a main-memory DBMS: 
fine-grained tuple-level logging lowers transaction rate since more data is recorded; 
coarse-grained transaction-level logging slows down 
failure-recovery phase as it incurs high computation overhead to replay the 
logs~\cite{malviya2014rethinking,zheng2014fast}. 
As we shall see, our proposed \system offers fast failure recovery 
without introducing additional runtime overhead.

\section{\systemcaption Overview}
\label{overview}

\mbox{\system} aims at providing fast failure recovery for modern main-memory DBMSs that
execute transactions as stored procedures~\cite{kallman2008h,stonebraker2007end,thomson2012calvin}.
A stored procedure is modeled as a \textit{parameterized transaction template} 
identified by a unique name that consists of a structured flow of database operations.
For simplicity,
we respectively abstract the \textit{read} and \textit{write} operations in a stored procedure as 
\texttt{var$\leftarrow$read(tbl, key)} and \texttt{write(tbl, key, val)}. 
Both operations search tuples in the table \texttt{tbl} using the candidate key called \texttt{key}. 
The read operation assigns the retrieved value to a local variable \texttt{var}, 
while the write operation updates the corresponding value to \texttt{val}. 
Insert and delete operations are treated as special write operations.
A client issues a request containing a procedure name and 
a list of arguments to initiate the execution of a \textit{procedure instance}, 
called a \textit{transaction}. 
The DBMS dispatches a request to a single \textit{worker} thread, 
which executes the initiated transaction to either commit or abort.

\system is designed for transaction-level logging~\cite{malviya2014rethinking}
that minimizes the runtime overhead for transaction processing.
The DBMS spawns a collection of \textit{logger} threads to 
continuously dump committed transactions to the secondary storage. 
To limit the log file size and facilitate parallel recovery,
the DBMS stores log entries into a sequence of files referred to as \textit{log batches}.
Each log entry records the stored procedure being invoked together with its input parameter values.
The entries in each log batch are strictly ordered according to the transaction commitment order.
The sequence of log batches are reloaded and processed in order during recovery.

Both the logging and log reloading can be performed in parallel, 
and we refer to \cref{implementation} for detailed discussions. 
In this paper, we focus on parallelizing the replay of the logs 
generated by transaction-level logging.

\begin{figure}[ht!]
\centering
\includegraphics[width=\columnwidth]{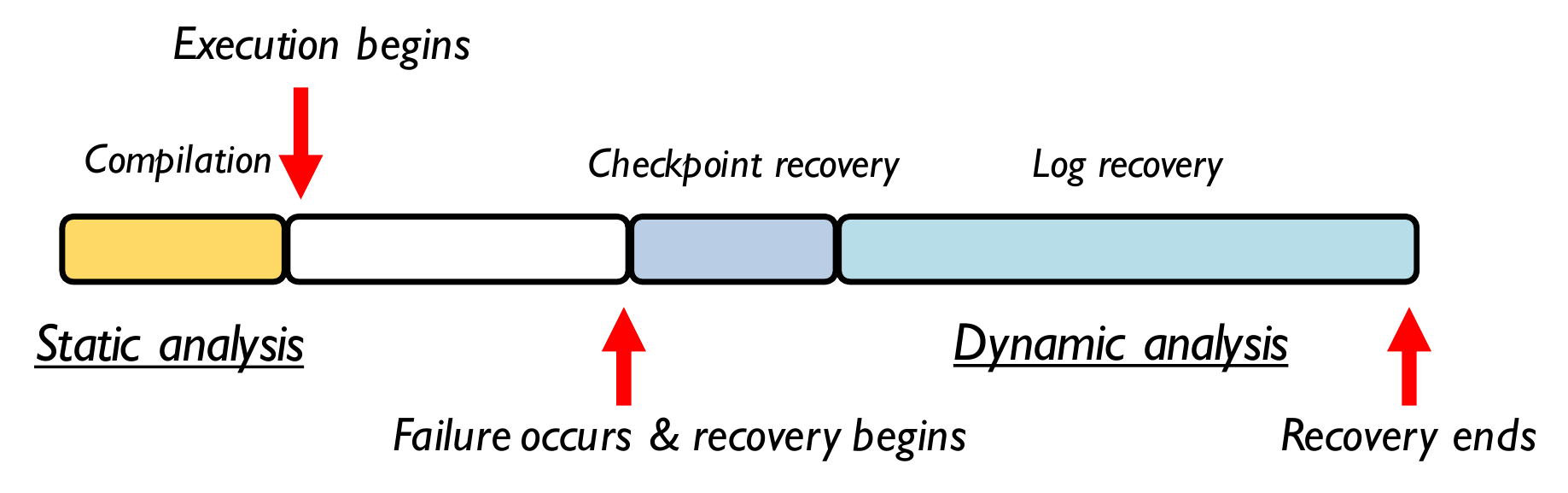}
\caption{Workflow of \system.}
\label{fig:overview}
\end{figure}

The workflow of \system is summarized in \cref{fig:overview}.
At compile time, 
\system performs a static analysis of the stored procedures to identify opportunities for parallel execution.
This analysis is performed in two stages.
In the first stage, each stored procedure is analyzed independently 
to identify the flow and data dependencies among its operations.
A flow dependency between two operations constrains the execution ordering between these operations,
while a data dependency between two operations indicates that 
these operations could potentially conflict
(i.e., one is reading and the other is writing the same tuple).
Based on the identified dependencies, 
the stored procedure is segmented into a maximal set of smaller pieces 
which are organized into a directed acyclic graph, referred to as a \textit{local dependency graph}.
This graph explicitly captures 
the possible parallelization opportunities as well as 
the execution ordering constraints among the pieces.
In the second stage, the local dependency graphs derived from the stored procedures
are integrated into a single dependency graph,
referred to as a \textit{global dependency graph}. This graph captures 
execution ordering among the different subsets of pieces from all the procedures.

During recovery, \system generates an execution schedule for each log batch using the 
global dependency graph.
A straightforward approach to replay the log batches
would be executing the schedules serially following the order of the log batches.
For each schedule, instantiations of the stored procedure pieces could be executed
in parallel following the execution ordering constraints derived from the global dependency graph.

To go beyond the execution parallelism obtained from static analysis,
\system further applies a dynamic analysis of the generated execution schedules 
to obtain a higher degree of parallelism in two ways.
First,
by exploiting the availability of the runtime procedure parameter values,
\system enables further intra-batch parallel executions.
Second, by applying a pipelined execution optimization, 
\system enables inter-batch parallel executions
where different log batches are  replayed in parallel.

In the following section, we discuss the design of \system in detail.

\section{\systemcaption Design}
\label{recovery}

\system achieves speedy failure recovery with a combination of static and dynamic analyses. 
In this section, we first show how \system leverages static analysis 
to extract flow and data dependencies out of predefined stored procedures 
at compile time (\cref{sec:staticanalysis}). 
We then explain how the static analysis can enable coarse-grained parallel recovery
(\cref{sec:schedules}).
After that, we discuss how dynamic analysis is used to achieve a
high degree of parallelism during recovery time (\cref{sec:recovery} and \cref{sec:runtime}).
We further elaborate how \system recovers
ad-hoc transactions without degrading the performance (\cref{sec:adhoc}). 

\subsection{Static Analysis}
\label{sec:staticanalysis}
\system performs static analysis at compile time to identify parallelization opportunities both within and across transactions.
This is captured through detecting the flow and data dependencies within each stored procedure and among different stored procedures.

\subsubsection{Intra-Procedure Analysis}
\label{sec:intraprocedure}
\system statically extracts operation dependencies from each stored procedure 
and constructs a \textit{local dependency graph} to characterize the execution ordering constraints 
among the operations in the procedure.
The corresponding algorithm is presented in \cref{sec:algorithms}.
Following classic program-analysis techniques~\cite{nielson1999principles,wu2016transaction,yan2016leveraging}, 
\system identifies \textit{flow dependencies} that capture two types of relations 
present in the structured flow of a program:
(1) define-use relation between two operations 
where the value returned by the preceding operation is used as input by the following operation;
(2) control relation between two operations 
where the output of the preceding operation determines 
whether the following operation should be executed.
Flow dependencies are irrelevant to operation type (e.g., read, write, insert, or delete),
and any operation can form flow dependencies with its preceding operations.

\begin{figure}[t!]
\centering
\includegraphics[width=\columnwidth]{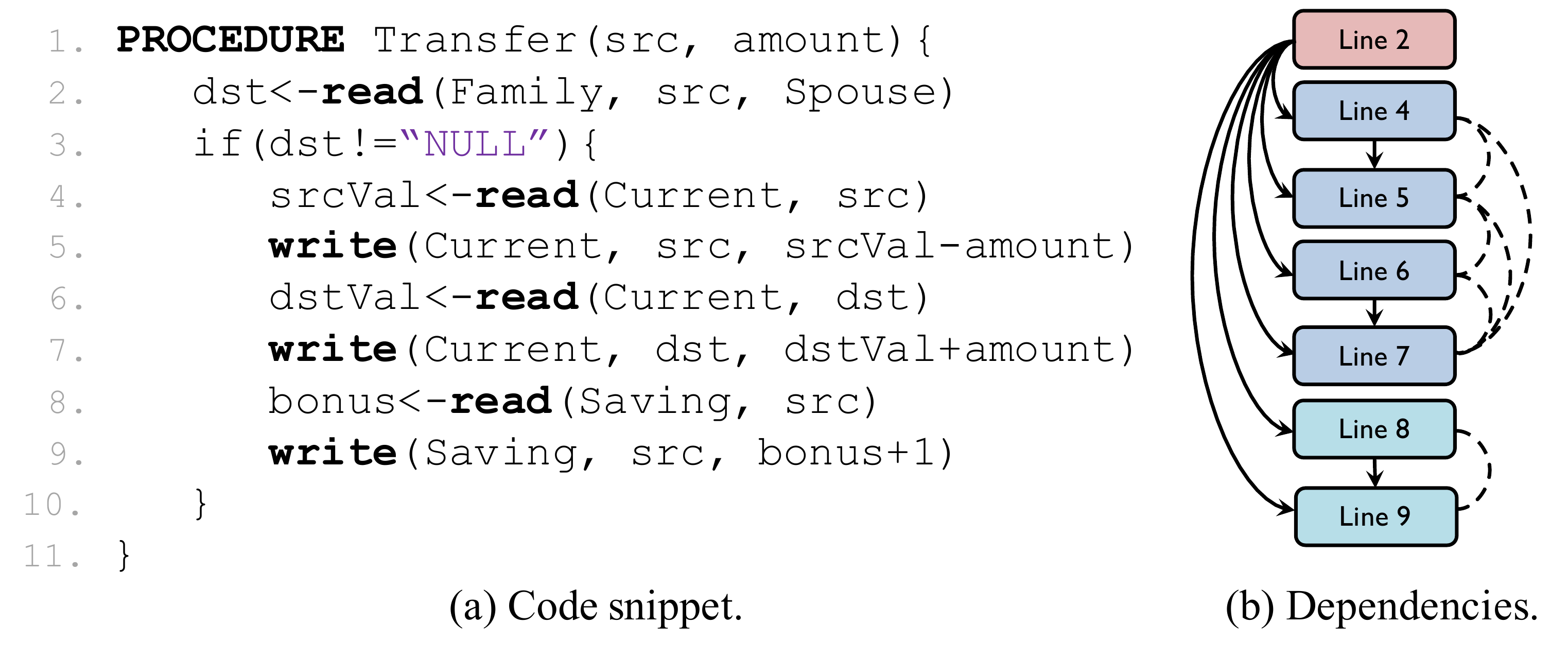}
\caption{Bank-transfer example. 
(a) Stored procedure.
(b) Flow (solid lines) and data (dashed lines) dependencies.}
\label{fig:intradependency}
\end{figure}

These two relations indicate the \textit{happen-before} properties among operations, 
and \textit{partially} restrict the execution ordering of the involved operations 
in a single stored procedure.
To illustrate these dependencies, 
consider the pseudocode in \cref{fig:intradependency}a resembling a bank-transfer example.
This stored procedure transfers an amount of money 
from a user's current account to her spouse's account, 
and adds one dollar bonus to the user's saving account. 
We say that the operation in Line~5 is \textit{flow-dependent} on that in Line~4, 
because the write operation uses the variable \texttt{srcVal} 
defined by the preceding read operation. 
Operations in Lines~4-9 are flow-dependent on the preceding read operation 
in Line~2 that generates the variable \texttt{dst}, 
which is placed on the decision-making statement in Line~3. 

Classic program-analysis techniques, 
including points-to analysis~\cite{steensgaard1996points} and 
control-dependency analysis~\cite{allen1970control}, 
can efficiently extract flow dependencies from stored procedures, 
and two flow-independent operations can be potentially executed in parallel 
at runtime~\cite{austin1992dynamic}. 
However, such analysis approaches ignore the data conflicts inherited in database accesses.
To address this problem, \system further identifies 
\textit{data dependencies} among operations to capture their potential ordering constraints.
Specifically, we say that two operations are \textit{data-dependent} 
if both operations access the same table and at least one of them is a modification operation.
Note that an insert or a delete operation can also form data-dependent relations
with other operations if both operate on the same table.
In the bank-transfer example, 
operations in Lines~4 and 5 are mutually data-dependent because they both access the \texttt{Current} table
and one of them updates the table.
All the dependencies in bank-transfer example are illustrated in \cref{fig:intradependency}b. 

The flow dependencies and data dependencies altogether
can constrain the execution ordering of the database operations in a single stored procedure. 
However, they differ in detailed semantics. 
A flow dependency captures \textit{must-happen-before} semantics, 
meaning that a certain operation can never be executed until 
its flow-dependent operations have finished execution. 
In contrast, a data dependency in fact only captures \textit{may-happen-before} semantics, 
and runtime information can be incorporated to relax this constraint, 
as will be elaborated in \cref{sec:recovery}.

\begin{figure}[t!]
\centering
\includegraphics[width=0.8\columnwidth]{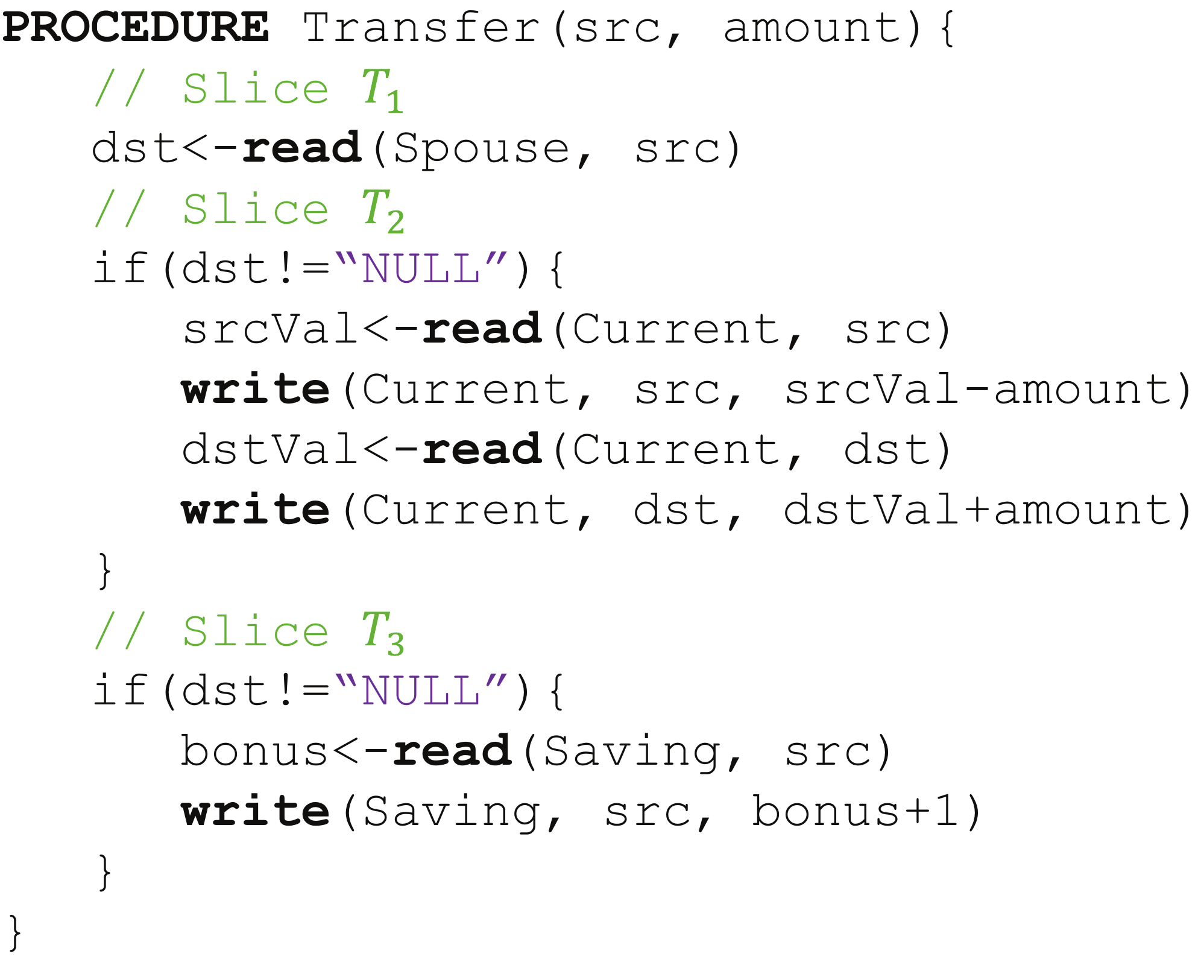}
\caption{Procedure slices in bank-transfer example.}
\label{fig:banktransferslice}
\end{figure}

Based on these dependencies, 
\system decomposes each procedure into a maximal collection of 
parameterized units called \textit{procedure slices} 
(or \textit{slices} for short) that satisfy the following two properties:
(1) each slice is a segment of a procedure program such that 
mutually data-dependent operations are contained in the same slice,
and
(2) whenever two operations $x$ and $y$ are in the same slice such that $y$ is flow-dependent on $x$, then any operation that is between $x$ and $y$ must also be contained in that slice.
\cref{fig:banktransferslice} shows the decomposition of the bank-transfer example into three slices (denoted by $T_1, T_2$, and $T_3$).

The set of slices decomposed from a stored procedure can be represented by a directed acyclic graph referred to as a
\textit{local dependency graph}.
The nodes in the graph correspond to the slices;
and there is a directed edge from one slice $s_i$ to another slice $s_j$
if there exists some operation $o_{j}$ in $s_j$ that is flow-dependent on some operation $o_{i}$ in $s_i$. 
The local dependency graph captures the execution order among the slices in the procedure as follows: 
for any two distinct slices $s_i$ and $s_j$ in the graph,
$s_i$ must be executed before $s_j$ if $s_i$ is an ancestor of $s_j$ in the graph;
otherwise, both slices could be executed in parallel
if $s_i$ is neither an ancestor nor a descendant of $s_j$ in the graph.

\cref{fig:intergraph}a illustrates the local dependency graph for the Transfer procedure in the bank-transfer example.
Observe that the operations in Lines~4-7 of \cref{fig:intradependency}a
are put into the same slice $T_2$ because these operations are mutually data-dependent.
Slices $T_2$ and $T_3$ are both flow-dependent on $T_1$
because the operations in $T_2$ and $T_3$ cannot be executed
until the variable \texttt{dst} has been assigned in the preceding read operation in Line 2.

\subsubsection{Inter-Procedure Analysis}
\label{sec:interprocedure}

\system further performs inter-procedure analysis to identify 
operation dependencies among the stored procedures. 
These dependencies are represented by a \textit{global dependency graph}
which is formed by integrating the local dependency graphs from all the stored procedures.
The detailed algorithm is presented in \cref{sec:algorithms}.

Before we formally define a global dependency graph,
we first extend the definition of data-dependent operations to data-dependent slices.
Given two procedure slices $s_i$ and $s_j$,
where $s_i$ and $s_j$ are slices from two distinct stored procedures,
we say that these slices are \textit{data-dependent}
if $s_i$ contains some operation $o_i$,
$s_j$ contains some operation $o_j$,
and both operations are data-dependent.

The global dependency graph $G$ for a set of stored procedures $P$ is a directed acyclic graph
where each node $v_i$ in $G$ represents a subset of procedure slices
from the local dependency graphs associated with $P$.
There is a directed edge from a node $v_i$ to another node $v_j$ in $G$
if $v_i$ contains some slice $s_i$, $v_j$ contains some slice $s_j$,
and both $s_i$ and $s_j$ are from the same stored procedure such that
$s_j$ is flow-dependent on $s_i$.
The nodes in $G$ satisfy the following four properties:
(1) each slice in $P$ must be contained in exactly one node in $G$;
(2) two slices that are data-dependent must be contained in the same node;
(3) if two nodes in $G$ are reachable from each other, these two nodes are merged into a single node;
and
(4) if a node contains two slices from the same stored procedure, these two slices are merged into a single slice.

For convenience, we refer to the set of slices associated with each node in $G$ as a \textit{block},
and we say that a block $B_j$ is \textit{dependent} on another block $B_i$ 
in $G$ if there is a directed edge from $B_i$ to $B_j$.

While a local dependency graph captures only the execution ordering constraints among slices from the same 
stored procedure,
a global dependency graph further captures the execution ordering constraints among slices from different stored
procedures.
Specifically,
for any two slices $s_i$ and $s_j$ in $G$,
where $s_i$ is contained in block $B_i$ and $s_j$ is contained in block $B_j$,
$s_i$ must be executed before $s_j$ if $B_i$ is an ancestor of $B_j$ in $G$;
otherwise, both slices could be executed in parallel
if $B_i$ is neither an ancestor nor a descendant of $B_j$ in $G$.

\begin{figure}[t!]
\centering
\includegraphics[width=0.8\columnwidth]{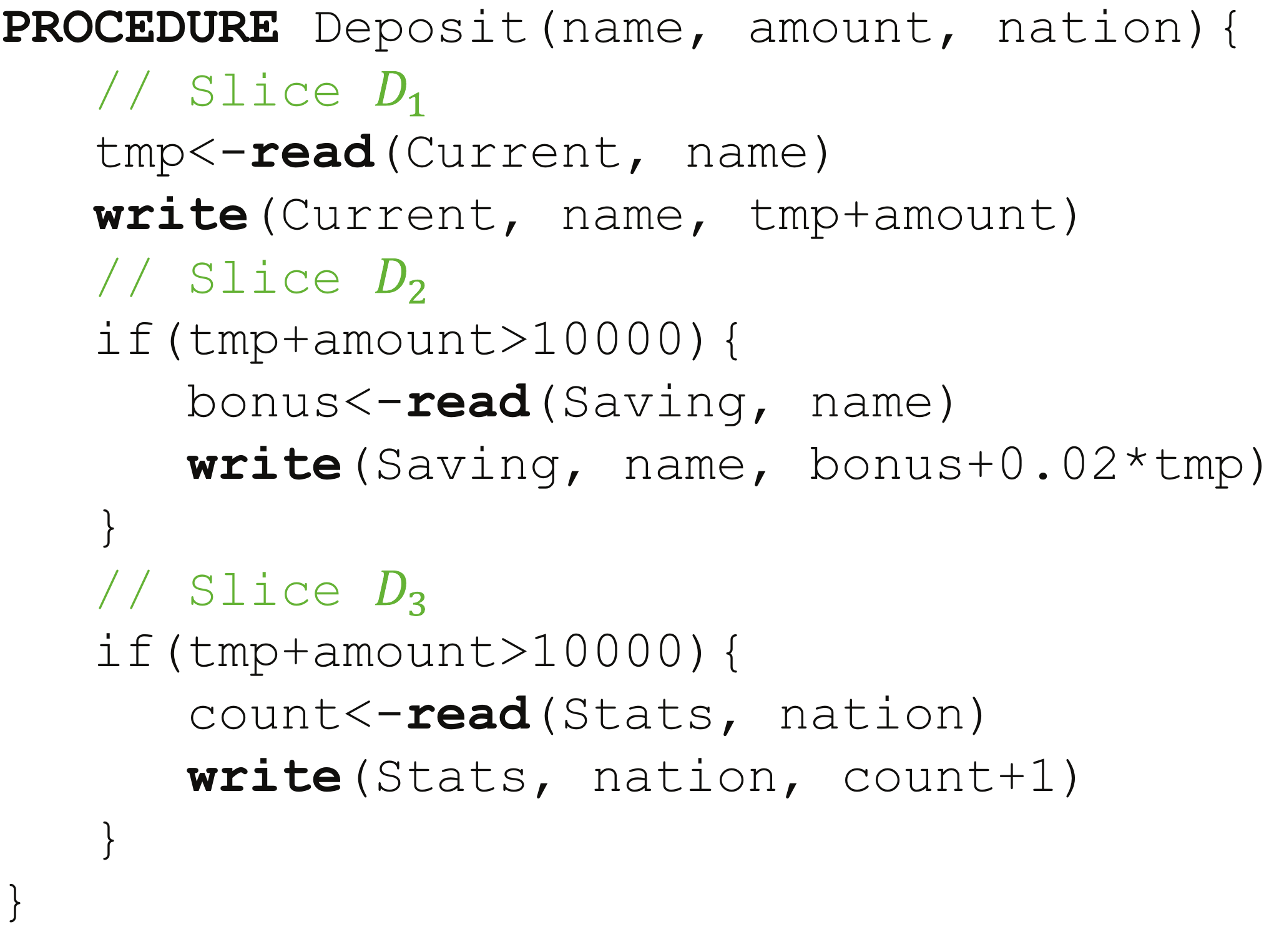}
\caption{Procedure slices in bank-deposit example.}
\label{fig:bankbonus}
\end{figure}

\begin{figure}[t!]
\centering
\includegraphics[width=0.9\columnwidth]{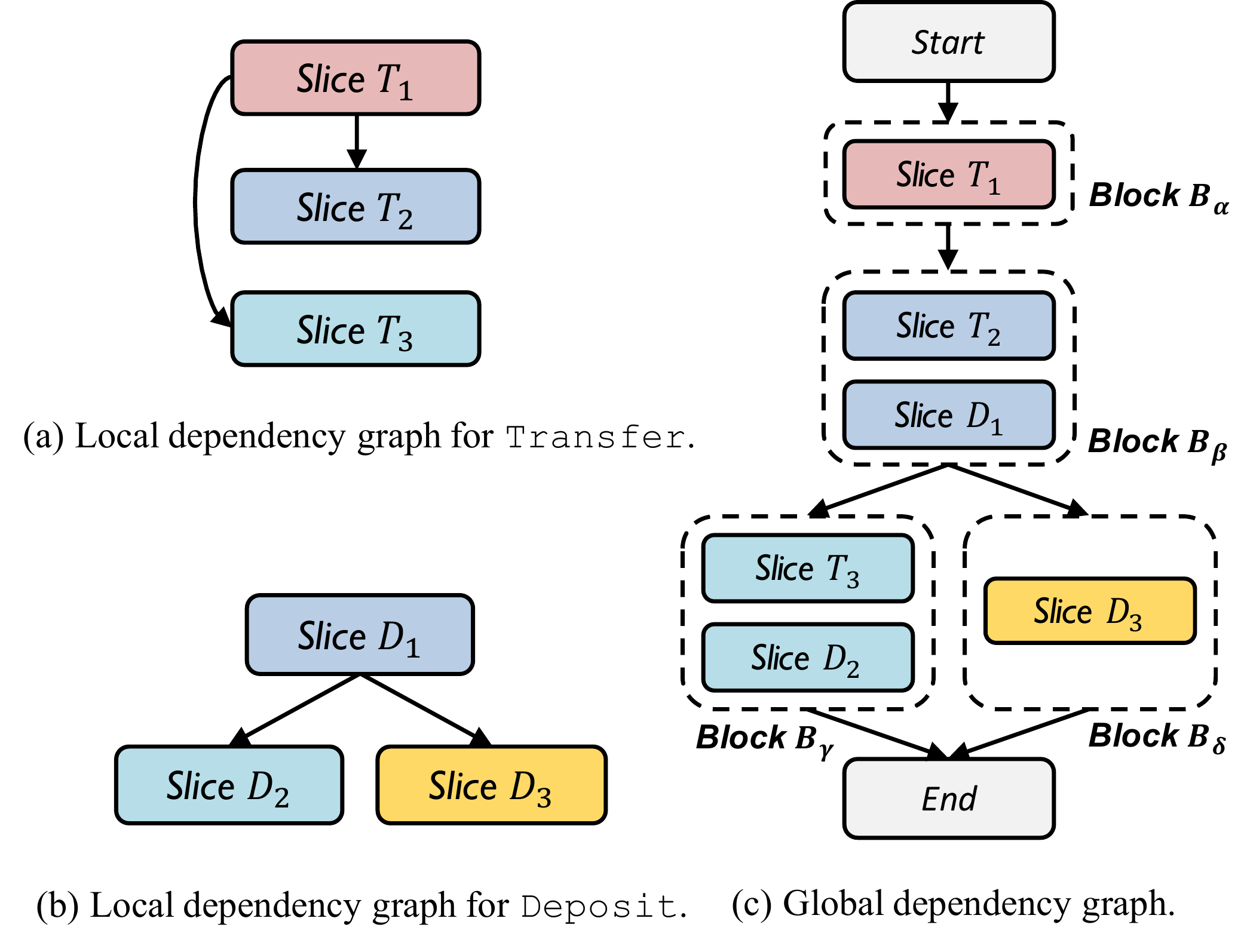}
\caption{
(a) and (b): Local dependency graphs for \texttt{Transfer} and \texttt{Deposit} procedures.
(c): Global dependency graph. 
Slices within the same dashed rectangle belong to the same block. 
Solid lines represent inter-block dependencies. 
}
\label{fig:intergraph}
\end{figure}

To give a concrete example, 
we introduce a second stored procedure,  named \texttt{Deposit},
that deposits an amount to some person's bank account, 
as shown in \cref{fig:bankbonus}. 
The local dependency graphs for these two procedures as well as the global dependency graph
for them are shown in \cref{fig:intergraph}. 
Observe that $T_2$ and $D_1$ are data-dependent slices residing in same block $B_\beta$.
For simplicity,
the dependency from $B_\alpha$ and $B_\gamma$ is omitted 
in the figure as it can be inferred from both the dependency from $B_\alpha$ to $B_\beta$
as well as the dependency from $B_\beta$ to $B_\gamma$.

\subsection{Recovery Execution Schedules}
\label{sec:schedules}

In this section, we explain how \system
could parallelize recovery from the log batches by exploiting the 
global dependency graph derived from static analysis.

\begin{figure}[t]
\centering
\includegraphics[width=\columnwidth]{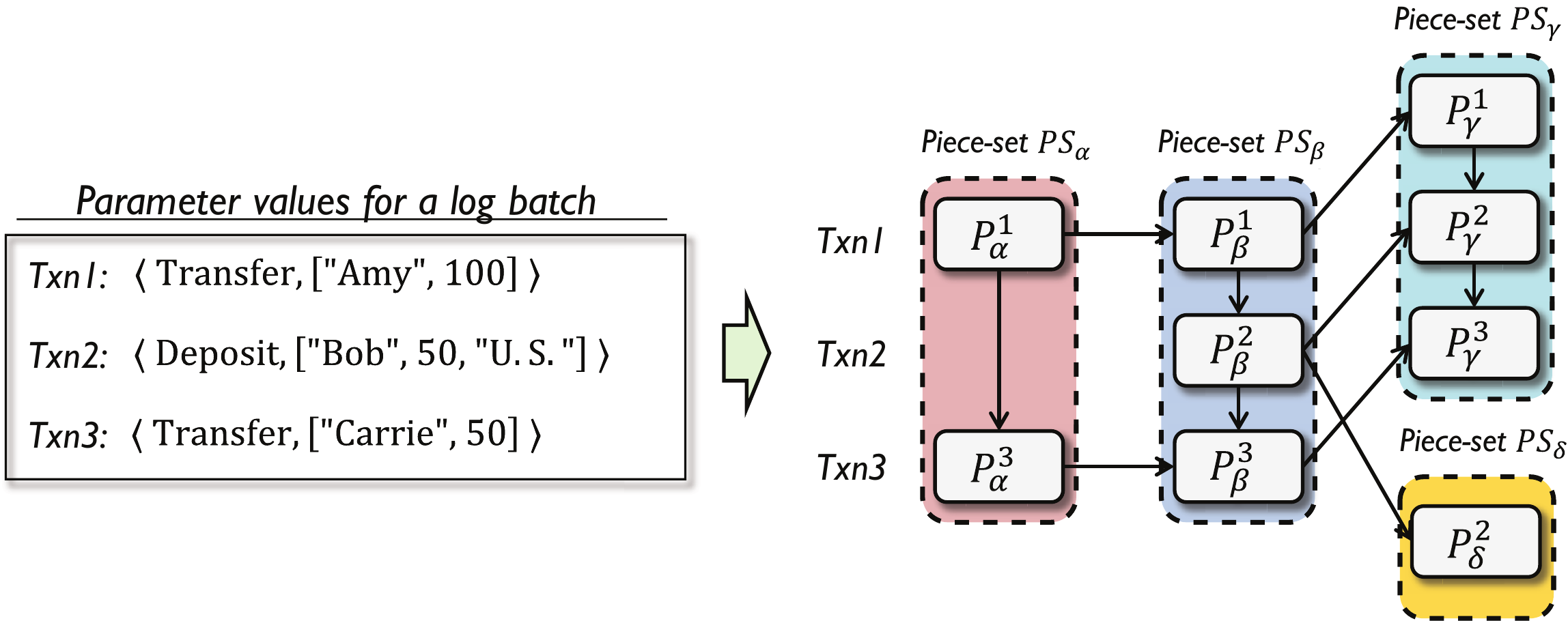}
\caption{Execution schedule for a log batch containing three transactions.}
\label{fig:schedule}
\end{figure}

During recovery, \system generates an execution schedule for each log batch using the 
global dependency graph (GDG).
We explain this process  using the example illustrated in
\cref{fig:schedule} for a simple log batch containing three transactions:
transactions \texttt{Txn1} and \texttt{Txn3}  invoke the \textit{Transfer} procedure,
while transaction \texttt{Txn2} invokes the \textit{Deposit} procedure. 

Recall that  \system applies a static analysis to segment each stored procedure into multiple slices 
to facilitate parallel execution.
Thus, each invocation of a stored procedure is actually executed in the form of a set of \textit{transaction pieces}
(or pieces for short)
corresponding to the slices for that procedure.
The execution schedule shown in \cref{fig:schedule} for the three transactions
is actually a directed acyclic graph of the transaction pieces
that are instantiated from the GDG in \cref{fig:intergraph}. 

Each transaction piece is denoted by $P_b^t$, 
where $t$ identifies the transaction order in the log batch
and $b$ identifies the block identifier in the GDG. 
For instance, \texttt{Txn2} is instantiated into three pieces:
$P_\beta^2$,
$P_\gamma^2$
and
$P_\delta^2$.
The directed edges among these pieces for a transaction reflect the dependencies
of their corresponding slices from the GDG.
The pieces from all three transactions are organized into four piece-sets
($PS_\alpha$, $PS_\beta$, $PS_\gamma$, and $PS_\delta$).
The pieces within the same piece-set correspond to slices in the same GDG block,
and these pieces are ordered (as indicated by the directed edges between them)
following the transaction order in the batch log.

We say that a piece $p$ is dependent on another piece $p'$ (or $p'$ is a dependent piece of $p$)
in an execution schedule  $ES$ if $p$ is reachable from $p'$ in $ES$.

Given an execution schedule for a log batch, 
the replay of the schedule during recovery must respect the dependencies among the pieces.
Specifically, a piece can be executed if all its dependent pieces have completed executions.
For example, for the execution schedule in
\cref{fig:schedule},
the piece $P_\gamma^2$ can be executed once its dependents ($P_\gamma^1$ and $P_\beta^2$)
have completed executions,
and the piece $P_\gamma^2$  
could be executed in parallel with both
$P_\delta^2$ and $P_\beta^3$.

\subsubsection{Efficient Coarse-Grained Parallelism}
\label{sec:intrabatch-cg}

While the above approach enables each log batch to be replayed with
some degree of fine-grained parallelism during recovery,
it could incur expensive coordination overhead when concurrent execution is enabled. 
This is because any transaction piece will need to initiate the execution of possibly multiple 
child pieces, and such initiation essentially requires accessing 
synchronization primitives for notifying concurrent threads.
As an example, the completion of piece $P_\beta^1$ will result in two primitive accesses 
for the initiation of $P_\beta^2$ and $P_\gamma^1$, 
while piece $P_\beta^2$ will lead to three coordination requests.

To reduce the coordination overhead involved in activating many piece executions,
\system instead handles the coordination at the level of piece-sets
by executing each piece-set with a single thread\footnote{As we shall see in \cref{sec:intrabatch-fg}, 
\system can parallelize the execution of a piece-set 
after extracting fine-grained intra-batch parallelism.}.
The completion of a piece-set is accompanied with one or more coordination requests, 
each of which initiates the execution of another piece-set.
By coordinating the executions at the granularity of piece-sets,
the execution output generated by each piece from $PS_\alpha$ are delivered together,
subsequently activating the execution of $PS_\beta$ with only a single coordination request.
For a large batch of transactions, this
approach can improve the system performance significantly, 
as we shall see in our extensive experimental study.


\subsection{Dynamic Analysis} 
\label{sec:recovery}

In this section, we explain how \system could further optimize the recovery process with a dynamic analysis
of the execution schedules\footnote{
The analysis is dynamic in the sense that it utilizes the runtime log record information in contrast to the static predefined stored procedure information
used by static analysis.
}.
Specifically, the performance improvement comes from two techniques.
First, by exploiting the availability of the runtime procedure parameter values,
\system enables further intra-batch parallel executions.
Second, by applying a pipelined execution optimization, 
\system enables inter-batch parallel executions
where different log batches are  replayed in parallel.

\subsubsection{Fine-Grained Intra-Batch Parallelism}
\label{sec:intrabatch-fg}

\begin{figure}[t!]
\centering
\includegraphics[width=\columnwidth]{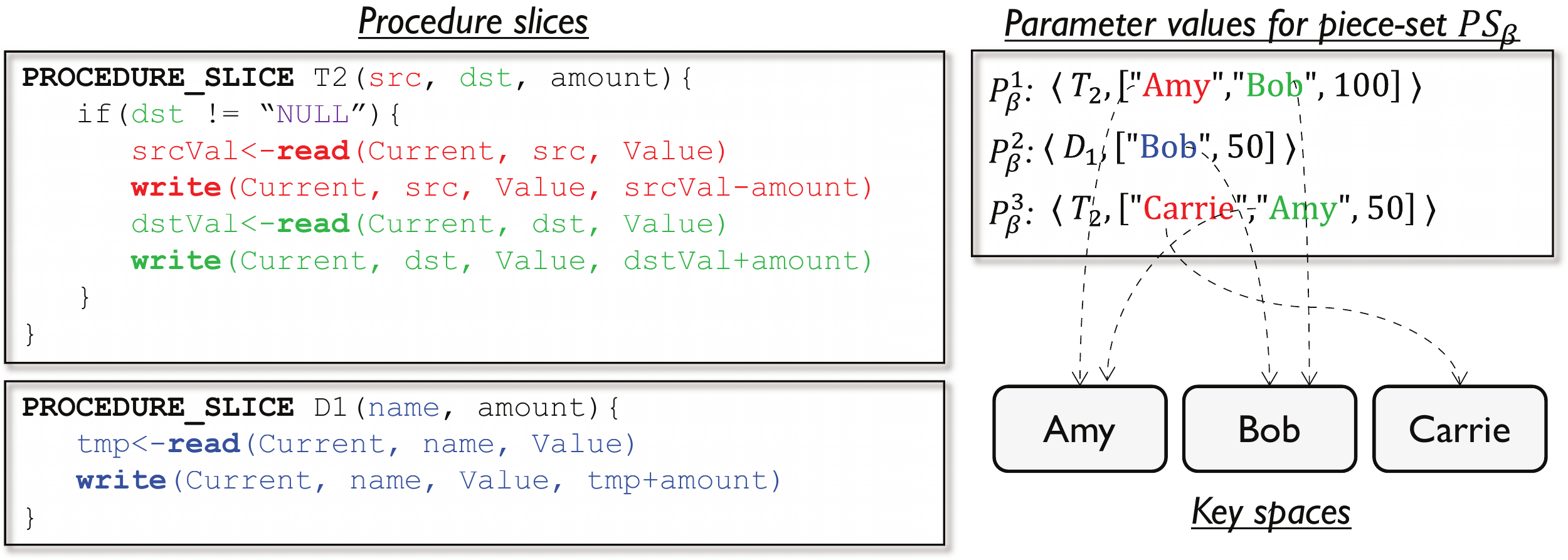}
\caption{Execution of piece-set $PS_{\beta}$ containing three transaction pieces.}
\label{fig:dispatch}
\end{figure}

Based on the discussion in \cref{sec:intrabatch-cg},
the transaction pieces within each piece-set will be executed following the transaction order in the log batch,
and the operations within each piece will also be executed serially.
As an example, consider the execution of the the piece-set $PS_\beta$ in \cref{fig:schedule},
where the three pieces in it are instantiated from the procedure slices $T_2$ and $D_1$ as 
shown in \cref{fig:dispatch}. 
The transaction pieces in $PS_{\beta}$ will be executed serially in the 
order $P_\beta^1$, $P_\beta^2$, and $P_\beta^3$;
and within a piece, for instance piece $P_\beta^1$ (which corresponds to slice $T_2$), 
the four operations inside will also be executed serially.
Such conservative serial executions are indeed inevitable if we are relying solely on the static analysis of the stored procedures.

However, given that the procedure/piece parameter values are actually available at runtime from 
both the log entries as well as the from those piece-sets that have already been replayed,
\system exploits such runtime information to further parallelize the execution of piece-sets.
Specifically, since the read and write sets of each transaction piece could be identified
from the piece's input arguments at replay time,
two operations in the same piece-set can be executed in parallel if they fall into 
different \textit{key spaces} (i.e., the two operations are not accessing the same tuple)
and there is no flow dependency between these operations.
Similarly, two pieces in a piece-set can be executed in parallel if their operations 
are not accessing any common tuple and there is no flow dependency between the piece-sets.

\begin{figure}[t!]
\centering
\includegraphics[width=0.75\columnwidth]{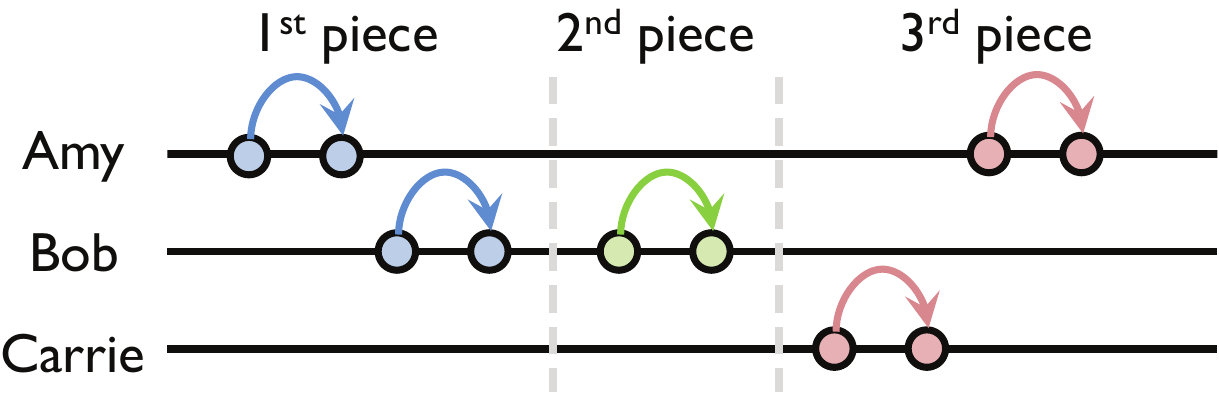}
\caption{Exploiting runtime information to identify accessed tuples in the execution of piece-set $PS_{\beta}$.
The flow dependencies (depicted by curved arrows) between  operations are known from static analysis.}
\label{fig:parallel1}
\end{figure}

Continuing with our example of the execution of the piece-set $PS_\beta$ in \cref{fig:dispatch},
the tuples accessed by each operation in these pieces can be identified by checking the input arguments. 
For example, the argument \texttt{Amy} in the piece $P_\beta^1$
identifies the accessed tuple for the first two operations listed in slice $T_2$, 
while \texttt{Bob} identifies the accessed tuple for the remaining two operations in $T_2$.
Similarly,  observe that
the tuple being accessed by the operations in $P_\beta^2$ is determined 
by the argument \texttt{Bob};
and 
the tuples being accessed by the operations in $P_\beta^3$ are determined 
by the arguments \texttt{Amy} and \texttt{Carrie}.
\cref{fig:parallel1} illustrates the tuples accessed by the operations in the execution of $PS_{\beta}$;
the flow dependencies shown are known from the static analysis.
Clearly, since the two tuples (with keys \texttt{Amy} and \texttt{Bob}) accessed by the two pairs of operations in $P_\beta^1$ (corresponding to slice $T_2$)
are distinct and there is no flow dependency between these pairs of operations,
these two pairs of operations can be safely executed in parallel without any coordination.
By a similar argument, the two pieces $P_\beta^2$ and $P_\beta^3$ can be executed in parallel
once the piece $P_\beta^1$ has completed execution.
It is important that the execution of $P_\beta^1$ be completed before 
starting $P_\beta^2$ and $P_\beta^3$ as 
the operations in $P_\beta^1$ conflict with those in each of
$P_\beta^2$ and $P_\beta^3$.

Observe that the flow dependencies shown for the execution of $PS_{\beta}$ in \cref{fig:parallel1} 
are due to what have been referred to as \textit{read-modify-write} access patterns~\cite{tu2013speedy}.
This access pattern involves two operations: the first operation reads a row and the second operation
updates the row read by the first operation.
As illustrated by the above discussion, if the read-modify-write patterns access different records, 
then the flow dependencies among these operations would not hinder their parallel executions.

Yet another commonly seen access pattern is what we call \textit{foreign-key} access pattern.
In a foreign-key pattern,
an operation reads a row $r_1$ from a table and then writes a related row $r_2$
in another table, where $r_1$ (or $r_2$) has a foreign key that refers to $r_2$ (or $r_1$).
Line~2 and Lines~4-5 in \cref{fig:intradependency} share this pattern\footnote{
This example is actually more sophisticated because Line~2 and Lines~4-5 fall into different slices. 
But we cannot prevent cases where operations in the same slice are flow-dependent.}, 
as the specific rows to be accessed in tables \texttt{Customer} and \texttt{Current} 
can be determined by \texttt{src}, meaning that these operations actually belong to the same key space. 

Both the read-modify-write and foreign-key access patterns are common in real-world applications.
In our analysis of fifteen well-known OLTP benchmarks~\cite{oltpbenchmark},
we observe that all the existing flow dependencies in these benchmarks 
are due to these two patterns.  
Moreover, our extensive experimental studies have also confirmed this observation. 
The prevalence of these two patterns indicates the  potential for parallel  operation executions.

\subsubsection{Inter-Batch Parallelism}
\label{sec:interbatch}

\begin{figure}[t!]
\centering
\includegraphics[width=\columnwidth]{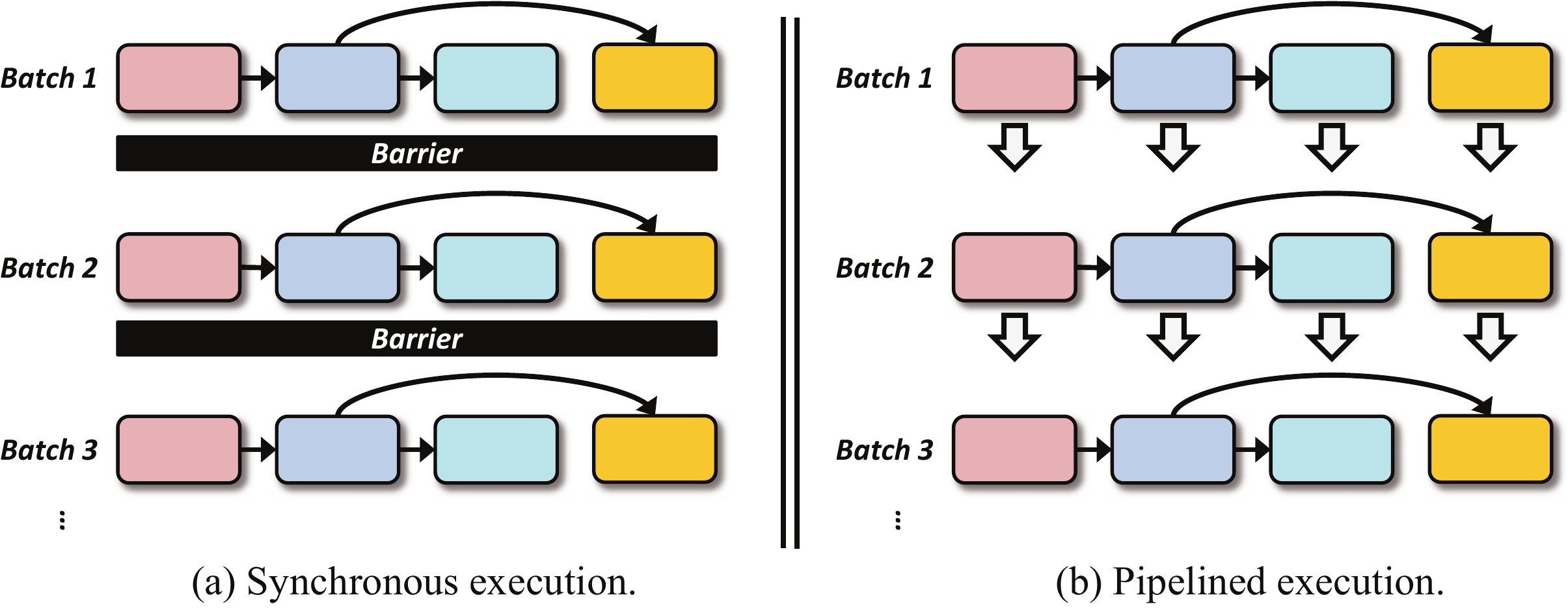}
\caption{
Synchronous execution vs pipelined execution for three log batches. 
Each rectangle represents a piece-set in an execution schedule.}\label{fig:pipeline}
\end{figure}

So far, our focus has been on intra-batch parallelism
to optimize the performance of executing an individual log batch schedule.
However, a DBMS usually need to recover tens of thousands of log batches
during the entire log recovery phase, as it is difficult to 
reload tens- or even hundreds-of-gigabyte of log data into DRAM at once.
By extracting purely intra-batch parallelism,
the DBMS has to execute log batches serially one after another,
and we refer to this execution mode as \textit{synchronous execution}.
As illustrated by the simple example in \cref{fig:pipeline}(a) showing
the execution of three log batches (which happen to have the same execution schedules),
such a serial execution requires synchronization barriers to coordinate the thread executions.
To enable inter-batch parallelism,
\system supports a \textit{pipelined execution} model
that enables a log batch to begin being replayed without having to wait for the replay of 
the preceding log batch to be entirely completed.
Specifically, a piece-set $P$ associated with a log batch $B$ could start execution
once its dependent piece-sets (w.r.t. $B$) and any piece-set in the same block as $P$ associated with its preceding
log batch have completed.

\subsection{Recovery Runtime}
\label{sec:runtime}


\mbox{\system} re-executes transactions as a pipeline of order-preserving data-flows,
which is facilitated by the combination of the static and dynamic analyses described above. 
Given the global dependency graph (GDG) generated at static-analysis stage,
\mbox{\system} estimates the workload distributions over the piece-sets of each procedure block
by counting the number of pieces at log file reloading time. 
Based on this distribution, \mbox{\system} assigns a fixed number of CPU cores 
in the machine to each block.
When a log batch is reloaded to main memory, \mbox{\system} generates an execution schedule
based on the GDG, where the instantiated piece-sets are one-to-one mapped to the blocks
in the GDG (see \mbox{\cref{sec:schedules}}).
\mbox{\system} thus can process each piece-set using the cores assigned to the
corresponding block, hence extracting coarse-grained recovery parallelism.
To enable finer-grained parallelism for recovery,
\mbox{\system} further dispatches operations inside a piece-set into 
different cores by exploiting the availability of the runtime 
parameter values (see \mbox{\cref{sec:intrabatch-fg}}).
This scheme allows \mbox{\system} to fully utilize computation resources
for processing a single log batch.
\mbox{\system} also exploits parallelisms across multiple log batches,
and this is achieved by pipelining the processing of different 
execution schedules (\mbox{\cref{sec:interbatch}}).


\begin{figure}[ht!]
\centering
\includegraphics[width=0.8\columnwidth]{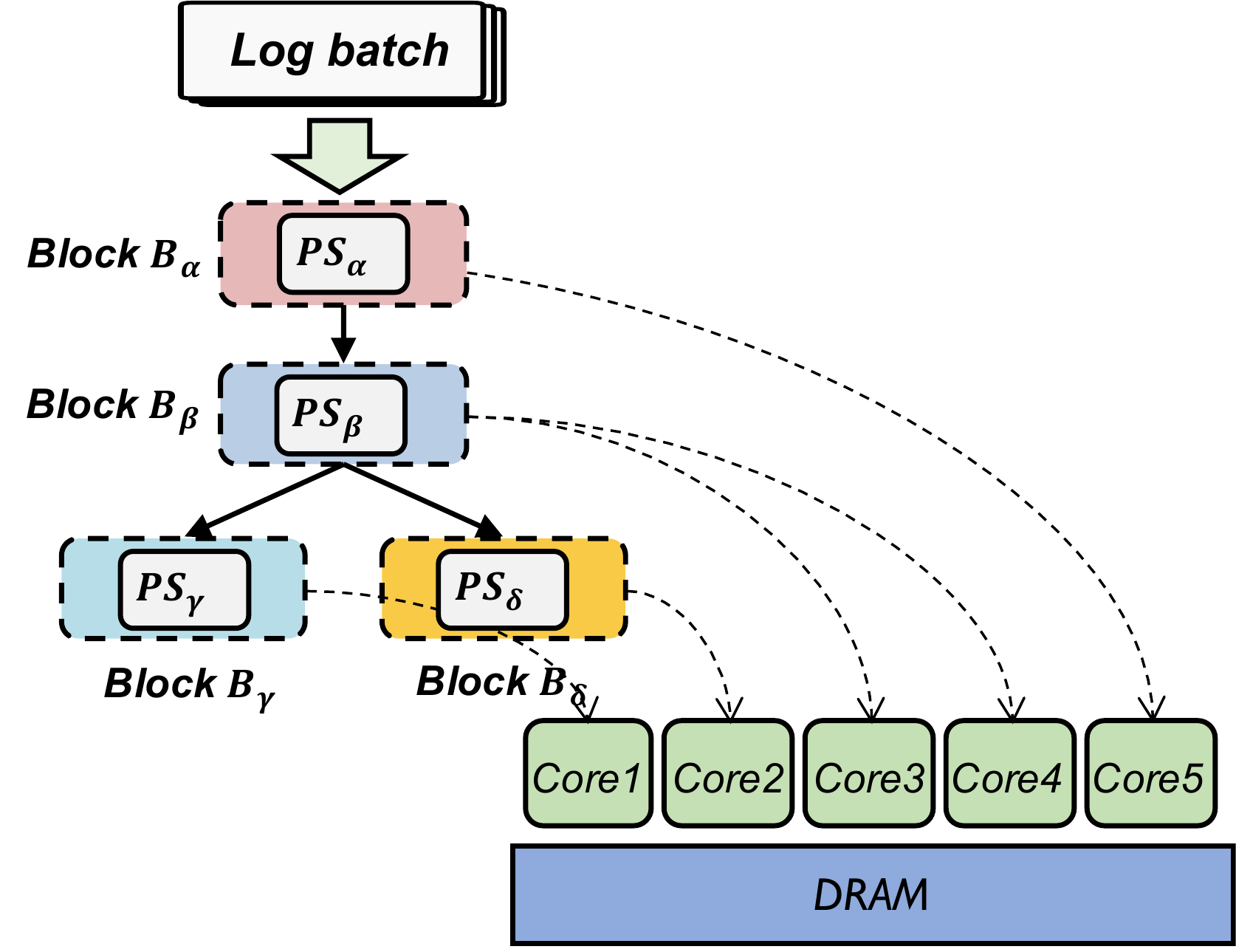}
\caption{Recovery runtime of \system. The workload distribution over the piece-sets of each block
($B_\alpha$, $B_\beta$, $B_\gamma$, and $B_\delta$) in the GDG is 20\%, 40\%, 20\%, and 20\%.}
\label{fig:deployment}
\end{figure}

\mbox{\cref{fig:deployment}} gives a concrete example of how \mbox{\system} 
performs database recovery for an application 
containing the \mbox{\texttt{Transfer}} and \mbox{\texttt{Deposit}} procedures. 
By estimating the workload distribution at log file reloading time,
\mbox{\system} assigns different number of cores to each block.
When processing a log batch, \mbox{\system} constructs an execution schedule and
splits the log batch into four piece-sets, namely $PS_\alpha$, $PS_\beta$, $PS_\gamma$, 
and $PS_\delta$. For a certain piece-set, for instance $PS_\beta$, \mbox{\system}
processes it using the two cores assigned to block $B_\beta$.
The operations within $PS_\beta$ are dispatched to these two cores using 
dynamic analysis. \mbox{\system} finishes processing this log batch once
all the four piece-sets have been recovered.
\mbox{\system}'s pipelined execution model further allows a log batch 
to be processed even if its preceding log batch is still under execution. 
\subsection{Ad-Hoc Transactions}
\label{sec:adhoc}

\system is designed for main-memory DBMSs that adopt 
command logging scheme for preserving database durability.
A known drawback of this logging scheme is that 
the execution behavior of a transaction containing 
nondeterministic operations (e.g., \texttt{SELECT * FROM FOO LIMIT 10}) 
cannot be precisely captured~\cite{malviya2014rethinking}.
Also, command logging does not naturally support transactions
that are not issued from stored procedures.
We refer to these transactions as ad-hoc transactions.
To support these transactions, a DBMS must additionally support 
conventional tuple-level logical logging to record every 
row-level modification of a transaction~\cite{malviya2014rethinking}.

The co-existence of both transaction-level and tuple-level logs
calls for a unified re-execution model that 
ensures the generality of our proposed recovery mechanism.
\system solves this problem by treating the replay of 
a transaction that is persisted using logical logging 
as the processing of a write-only transaction.
With the full knowledge of a transaction's write set, 
high degree of parallelism is easily extracted, 
as each write operation can be dispatched 
to the corresponding piece-subset of a certain block 
through dynamic analysis described in \cref{sec:recovery}.
Note that the replay of the tuple-level logs 
produced by ad-hoc transactions must still 
follow the strict re-execution order captured in the log batches.
As such, \system's solution enables 
the unification of recovery for 
transaction-level logging and tuple-level logging.

One extreme case for \system is that all the transactions processed
by the DBMS are ad-hoc transactions.
In this case, \system works essentially the same as a pure logical log
recovery scheme.
However, compared to existing solution~\cite{zheng2014fast},
\system does not need to acquire any latch during the log replay,
and hence, when multiple threads are utilized,
it yields much higher performance than existing tuple-level log recovery
schemes that employ latches during recovery.
This is confirmed by the experiment results shown in \cref{evaluations}.
\section{Discussion}
\label{discussion}

While \mbox{\system} provides performance benefits for transaction-level 
logging-and-recovery mechanisms, it has several limitations.

Foremost is that \mbox{\system} relies on the use of stored procedures.
Despite the fact that most DBMSs provide support for stored procedures, 
many application developers still prefer using dynamic SQL to query databases for reducing
the coding complexity.
Although this limitation can restrict the use of \mbox{\system},
an increasing number of performance-critical applications
such as on-line trading and Internet-of-Things (IoT) processing 
have already adopted stored procedures to avoid the round-trip communication cost.
\mbox{\system} is applicable for these scenarios without any modifications.

Second, \mbox{\system}'s static analysis requires the stored procedures to be deterministic queries
with read and write sets that can be easily computed.
Furthermore, it remains a challenging problem for \mbox{\system} to support nested transactions 
or transactions containing complex logic. 
As mentioned in \mbox{\cref{sec:adhoc}}, to address this problem, a DBMS has to resort to conventional
tuple-level logging for persisting every row-level modification of a transaction.

\section{Evaluation}
\label{evaluations}

In this section, we evaluate the effectiveness of \system, by seeking to 
answer the following key questions: 

\begin{itemize}[itemsep=1pt]
\item[1.] Does \system incur a significant logging overhead for transaction processing? 
\item[2.] Can \system achieve a high degree of parallelism during failure recovery?
\item[3.] How does each proposed mechanism contribute to the performance of \system?
\end{itemize}

We implemented \system in \database, a fully fledged main-memory DBMS 
optimized for high performance transaction processing. 
\database uses a B-tree style data structure for database indexes,
and it adopts multi-versioning for higher level of concurrency~\cite{wu2017an}.
In addition to \system, 
we also implemented the state-of-the-art 
tuple-level (both physical and logical) and transaction-level 
logging-and-recovery schemes in \database.
In our implementation, we have optimized the tuple-level logging-and-recovery
schemes by leveraging multi-versioning.
However, \system does not exploit any characteristics
of multi-versioning, as the design of \system makes no assumption
about the data layout, and it is general enough to be directly applicable for single-version DBMSs.
We present the implementation details in \cref{implementation}.

We performed all the experiments on a single machine running 
Ubuntu 14.04 with four 10-core Intel Xeon Processor E7-4820 clocked at 1.9 GHz, 
yielding a total of 40 physical cores. 
Each core owns a private 32 KB L1 cache 
and a private 256 KB L2 cache. 
Every 10 cores share a 25 MB L3 cache and a 32 GB local DRAM. 
The machine has 
two 512 GB SSDs with maximum sequential read and sequential write throughput of 550 and 520 MB/s respectively.

Throughout our experiments, 
we evaluated the DBMS performance using two well-known 
benchmarks~\cite{difallah2013oltp}, namely, TPC-C and Smallbank.
The global dependency graph for TPC-C is presented in \cref{sec:tpcc}.
Except for \mbox{\cref{figure:exp:logging_timeline_1ssd}}, which reports the logging performance using a single SSD,
all the other experiment results presented in this section adopt two SSDs, 
each assigned with a single logging thread and a single checkpointing thread~\cite{zheng2014fast}.


\subsection{Logging}

In this section, we investigate how different logging schemes 
influence the performance of transaction processing.
We first measure the runtime overhead incurred by different 
logging schemes, and then evaluate how ad-hoc transactions affect the performance of transaction-level logging scheme.
Our experiment results demonstrate the effectiveness of the transaction-level logging scheme. 

\subsubsection{Logging Overhead}

We begin our experiments by evaluating the runtime overhead incurred by 
each logging scheme when processing transactions in the TPC-C benchmark.
Similar trends were observed for the Smallbank benchmark.
We set the number of warehouses to 200 and the database size is approximately 20 GB\footnote{
Note that the database size measures only the storage space for tuples; 
the total storage space occupied by the tuples and other auxiliary structures (e.g., indexes, lock tables)
is about 70 GB.}.
Due to the memory limit of our experiment machine, we disabled 
the insert operations in the original benchmark so that the database size
will not grow without bound.
We configure \database to use 32 threads for transaction executions,
2 threads for logging, and 2 threads for checkpointing.
We further configure \database to perform checkpointing every 200 seconds.

\begin{figure}[t!]
    \centering
    \fbox{
    \includegraphics[width=0.9\columnwidth]
        {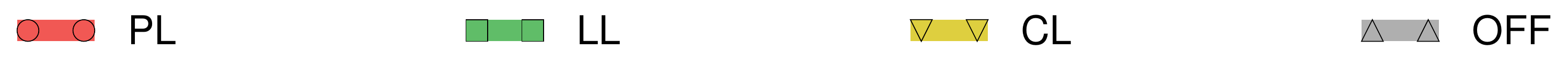}
    }
    \subfloat[With one SSD.]{
        \includegraphics[width=0.49\columnwidth]
            {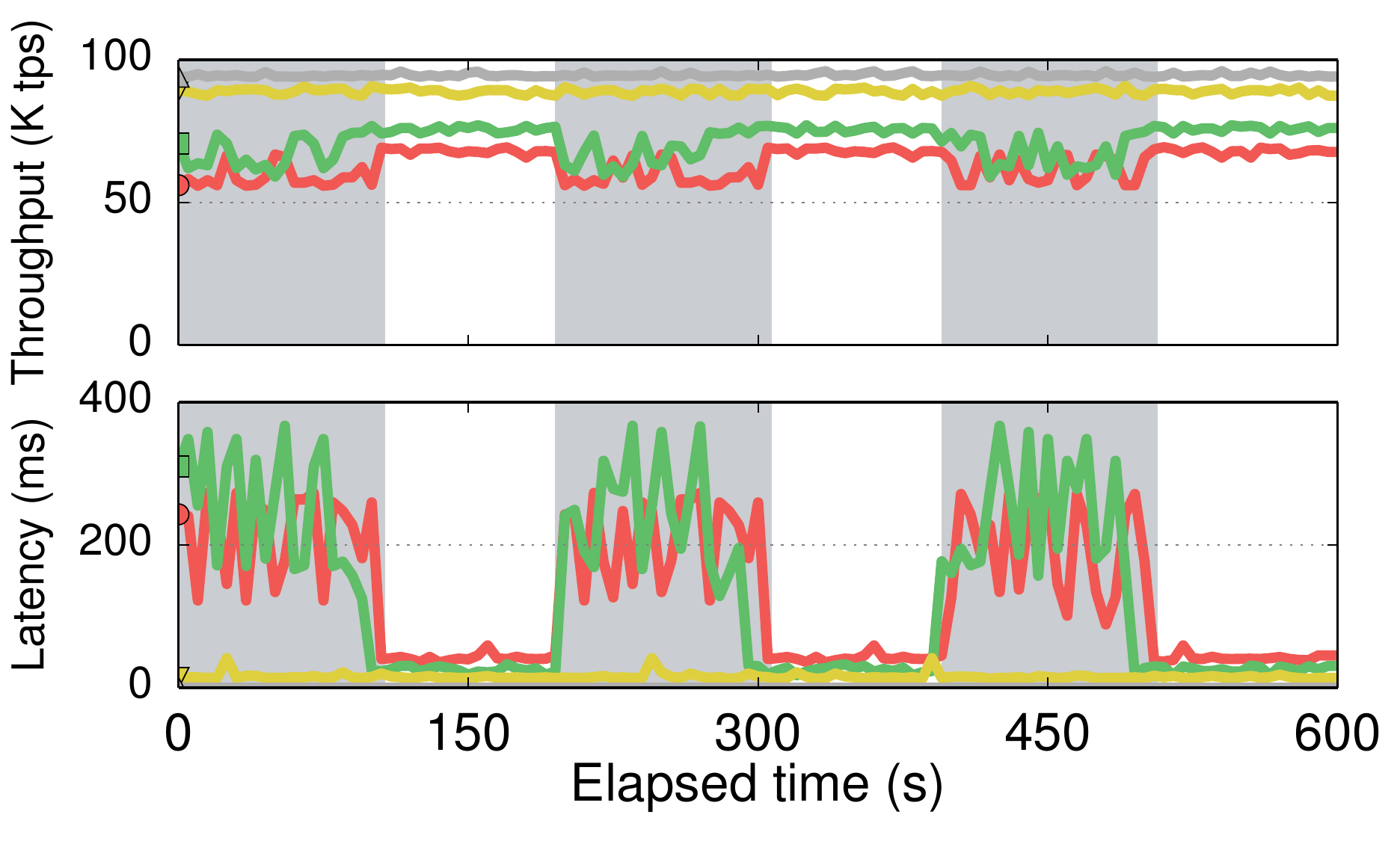}
    \label{figure:exp:logging_timeline_1ssd}
    }
    \subfloat[With two SSDs.]{
        \includegraphics[width=0.49\columnwidth]
            {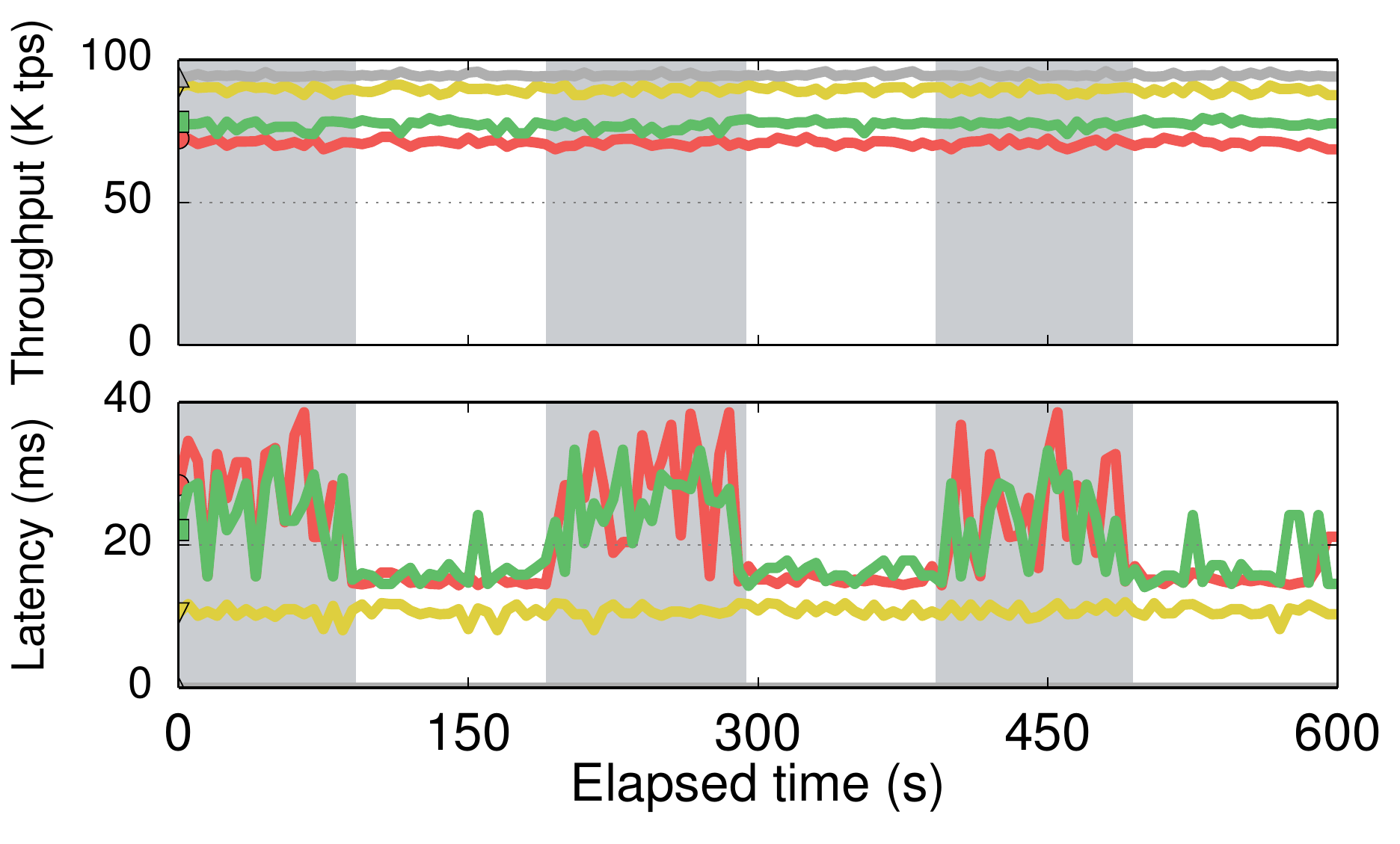}
    \label{figure:exp:logging_timeline_2ssd}
    }
    \caption{
        Throughput and latency comparisons during transaction processing.
        PL, LL, and CL
        stand for physical logging, logical logging, and command logging, respectively.
    }
    \label{figure:exp:logging_timeline}
\end{figure}

\begin{table}[t!]
    \centering
    \small
    \begin{tabular}{c|c|c|c|c|c|c|c|c}
    \hline
    & \multicolumn{3}{c|}{Throughput (K tps) } 
    & \multicolumn{3}{c|}{Log size (GB/min) } 
    & \multicolumn{2}{c}{Log size ratio} \\\hhline{~--------}
    & PL & LL & CL & PL & LL & CL & PL/CL & LL/CL \\\hline\hline
    \scriptsize TPC-C & 71 & 74 & 93 & 13.7 & 12.9 & 1.2 & 11.4 & 10.8 \\
    \scriptsize Smallbank & 503 & 564 & 595 & 1.6 & 1.2 & 1.3 & 1.23 & 0.92 \\\hline
    \end{tabular}\vspace{5pt} 
    \caption{Log size comparison.}
    \label{table:log_size}
\end{table}

\cref{figure:exp:logging_timeline} shows the throughput and the latency of the DBMS for the TPC-C benchmark a 10-minute duration.
Intervals during which the checkpointing threads are running are shown in gray.
With both logging and checkpointing disabled (denoted as OFF), 
the DBMS achieves a stable transaction processing throughput of around 95 K tps. 
However, the first 100-second trace in \mbox{\cref{figure:exp:logging_timeline_1ssd}} depicts that,
using one SSD, the throughput of the DBMS can drop by \mbox{$\sim$25\%} 
when both checkpointing and 
tuple-level logging, namely physical logging (denoted as PL) and logical
logging (denoted as LL), are enabled. 
When the DBMS finished performing checkpointing, the throughput rises to around 76 K tps
(see the throughput of LL from 100 to 200 seconds),
but this number is still 20\% lower than the case where recovery schemes in the DBMS 
are fully disabled.
Compared to tuple-level logging schemes,
the runtime overhead incurred by transaction-level logging, or command logging (denoted as CL),
is negligible.
Specifically, the throughput reduction caused by CL is under 6\%
even when checkpointing threads were running.

Tuple-level logging schemes also caused a significant increase in transaction latency.
As \cref{figure:exp:logging_timeline_1ssd} shows, 
there are high latency spikes when checkpointing threads were running.
In the worst case, the latency can go beyond 300 milliseconds, which is
intolerable for modern OLTP applications. 
To mitigate this problem, a practical solution
is to equip the machine with more storage devices.

\mbox{\cref{figure:exp:logging_timeline_2ssd}} shows the transaction throughput and latency 
achieved when persisting checkpoints and logs to two separate SSDs.
The result shows that adding more SSDs can effectively minimize the drop in throughput and significantly reduce the latency of tuple-level logging. 
However, tuple-level logging still incurs \mbox{$\sim$20\%} of throughput degradation, and its latency is at least twice higher than that of transaction-level logging.
These results demonstrate while the performance of tuple-level logging could be improved with additional storage devices, transaction-level logging still outperforms tuple-level logging.

The major factor that causes the results shown above is that tuple-level logging
schemes usually generate much more log records than transaction-level logging, and the SSD bandwidth can be easily saturated
when supporting high throughput transaction processing.
As shown in \cref{table:log_size},
the log size generated by logical logging in the TPC-C benchmark can be 10.8X
larger than that generated by command logging.
Physical logging yields an even larger log size because it must record the locations of the
old and new versions of every modified tuple. 
In the Smallbank benchmark, while the log size 
generated by the different logging schemes are similar,
command logging still yields comparatively better performance than the other schemes.
This is because log data serialization in physical and logical
logging schemes requires the DBMS to iterate a transaction's write set 
and serialize every attribute of each modified tuple
into contiguous memory space. This process leads to higher
overhead than that in command logging.
\mbox{\cref{sec:logging-performance}} presents additional analysis of the impact of SSD bandwidth and \mbox{\texttt{fsync}} operations on the performance of the different logging schemes.

\begin{figure}[t!]
    \centering
    \fbox{
    \includegraphics[width=0.9\columnwidth]
        {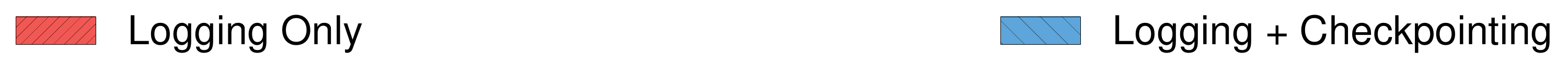}
    }
    \subfloat[Throughput.]{
        \includegraphics[width=0.49\columnwidth]
            {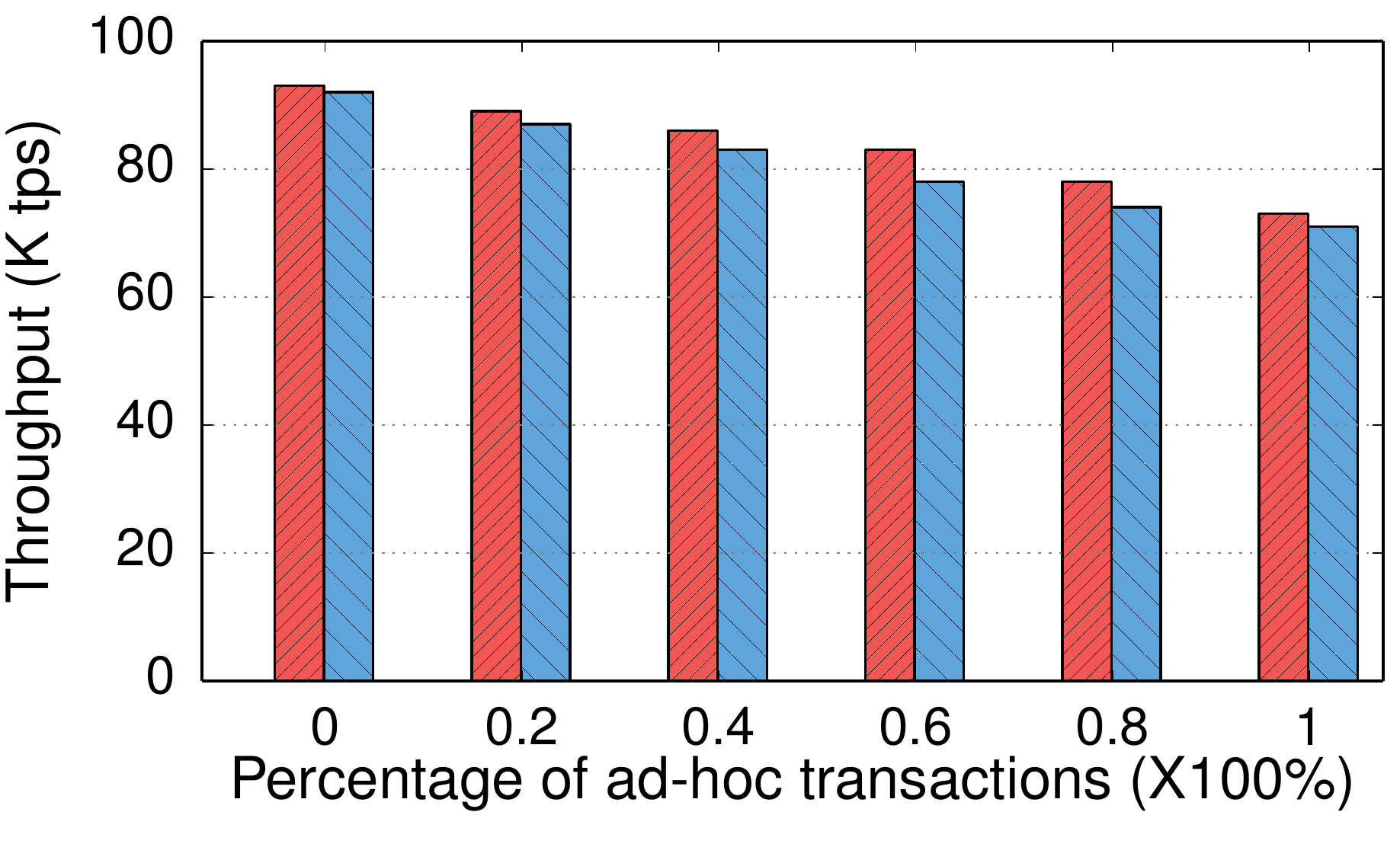}
        \label{figure:exp:logging_throughput_adhoc}
    }
    \subfloat[Latency.]{
        \includegraphics[width=0.49\columnwidth]
            {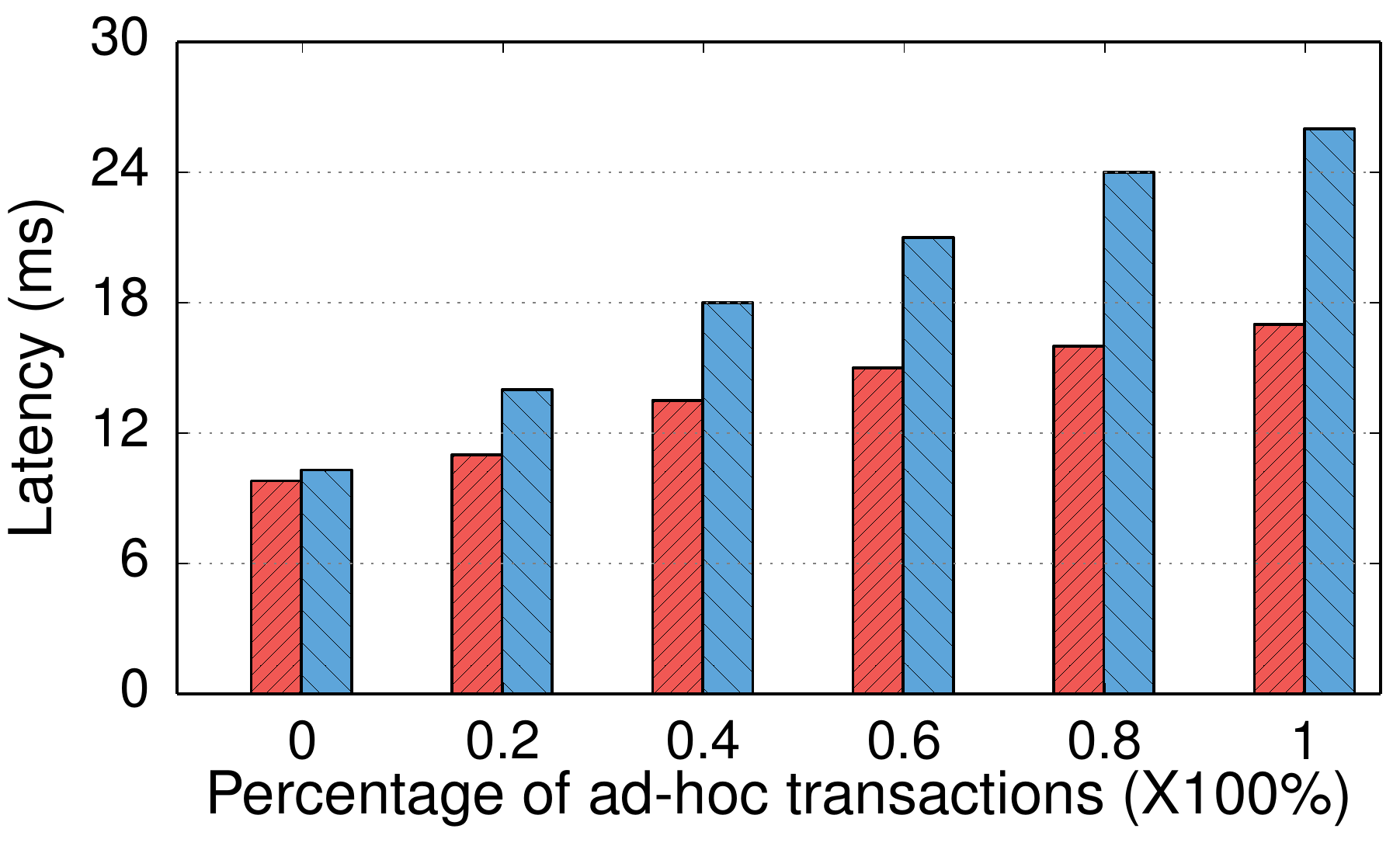}
        \label{figure:exp:logging_latency_adhoc}
    }
    \caption{
        Logging with ad-hoc transactions. 
    }
    \label{figure:exp:logging_adhoc}
\end{figure}

\subsubsection{Ad-Hoc Transactions}
As discussed in \cref{sec:adhoc}, 
the logging of ad-hoc transactions incurs additional overhead as the DBMS needs to log row-level modifications.
In this section, we evaluate the logging overhead for ad-hoc transactions using the TPC-C benchmark.
Similar trends were observed for Smallbank benchmark.
In our experiment, we randomly tag some transactions as ad-hoc transactions.
As shown in \cref{figure:exp:logging_throughput_adhoc}, 
the transaction throughput achieved by the DBMS drops almost linearly 
with the increase of the percentage of ad-hoc transactions.
\cref{figure:exp:logging_latency_adhoc} further shows
that the transaction latency increases significantly with the increase in percentage of ad-hoc transactions
especially when checkpointing is performed along with logging.
When 100\% of the transactions are ad-hoc, the performance degrades significantly as
the DBMS essentially ends up performing pure logical logging.
Based on these results, we confirm that the overhead incurred by command logging is no higher than that incurred by logical logging. 


\subsection{Recovery}

This section evaluates the performance of \system for database recovery.
Our evaluation covers the following schemes:
\begin{itemize}
\item 
\textbf{PLR:} 
This is the physical log recovery scheme that is widely implemented
in conventional disk-based DBMSs.
It first reloads and replays
the logs to restore tables with committed updates using multiple threads. 
After that, it rebuilds all the indexes in parallel. 
It adopts last-writer-wins rule to reduce log recovery time.
A recovery thread must first acquire a latch on any tuple that is to be modified.
The recovered database state is multi-versioned.
\item
\textbf{LLR:} This is the state-of-the-art logical log recovery scheme
proposed in SiloR~\cite{zheng2014fast}. 
It reconstructs the lost database records and indexes at the same time. 
While the original scheme was designed for single-version DBMSs,
we have optimized this scheme by exploiting multi-versioning to enable two recovery threads 
to restore different versions of the same tuple in parallel.
To ensure that all new tuple versions are appended correctly to the appropriate version chains,
latches are acquired by the recovery threads on the tuples being modified.
The recovered database state is multi-versioned. 
\item
\textbf{LLR-P:} This is the parallel logical log recovery scheme adapted from \system (see \cref{sec:adhoc}).
It treats the restoration of each transaction log entry as the replay of
a write-only transaction. 
During the log replay, it shuffles the write operations according to the table ID and primary key.  
After that, it reinstalls these operations in a latch-free manner.
The recovered database state is single-versioned. 
\item
\textbf{CLR:} This is the conventional approach for command log recovery. It 
reloads log files into memory in parallel and then re-executes the lost committed transactions in sequence using a single thread.
The recovered database state is single-versioned.
\item
\textbf{CLR-P:} This is the parallel command log recovery scheme (\system) described in this paper.
The recovered database state is single-versioned. 
\end{itemize}

The entire database recovery process
operates in two stages: 
(1) checkpoint recovery, which restores the database to the 
transactionally-consistent state at the last checkpoint;
and
(2) log recovery, which reinstalls the effects made by all the lost committed transactions. 
We study these two stages separately, and then evaluate 
the overall performance of the entire database recovery process.
Finally, we study the effect of ad-hoc transactions.

\subsubsection{Checkpoint Recovery}

\begin{figure}[t!]
    \centering
    \fbox{
    \includegraphics[width=0.9\columnwidth]
        {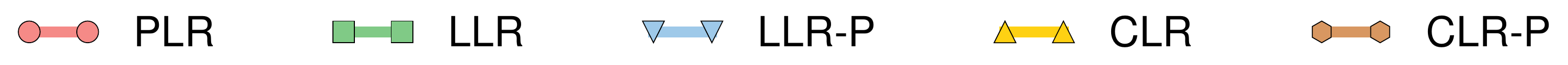}
    }
    \subfloat[Pure checkpoint file reloading.]{
        \includegraphics[width=0.49\columnwidth]
            {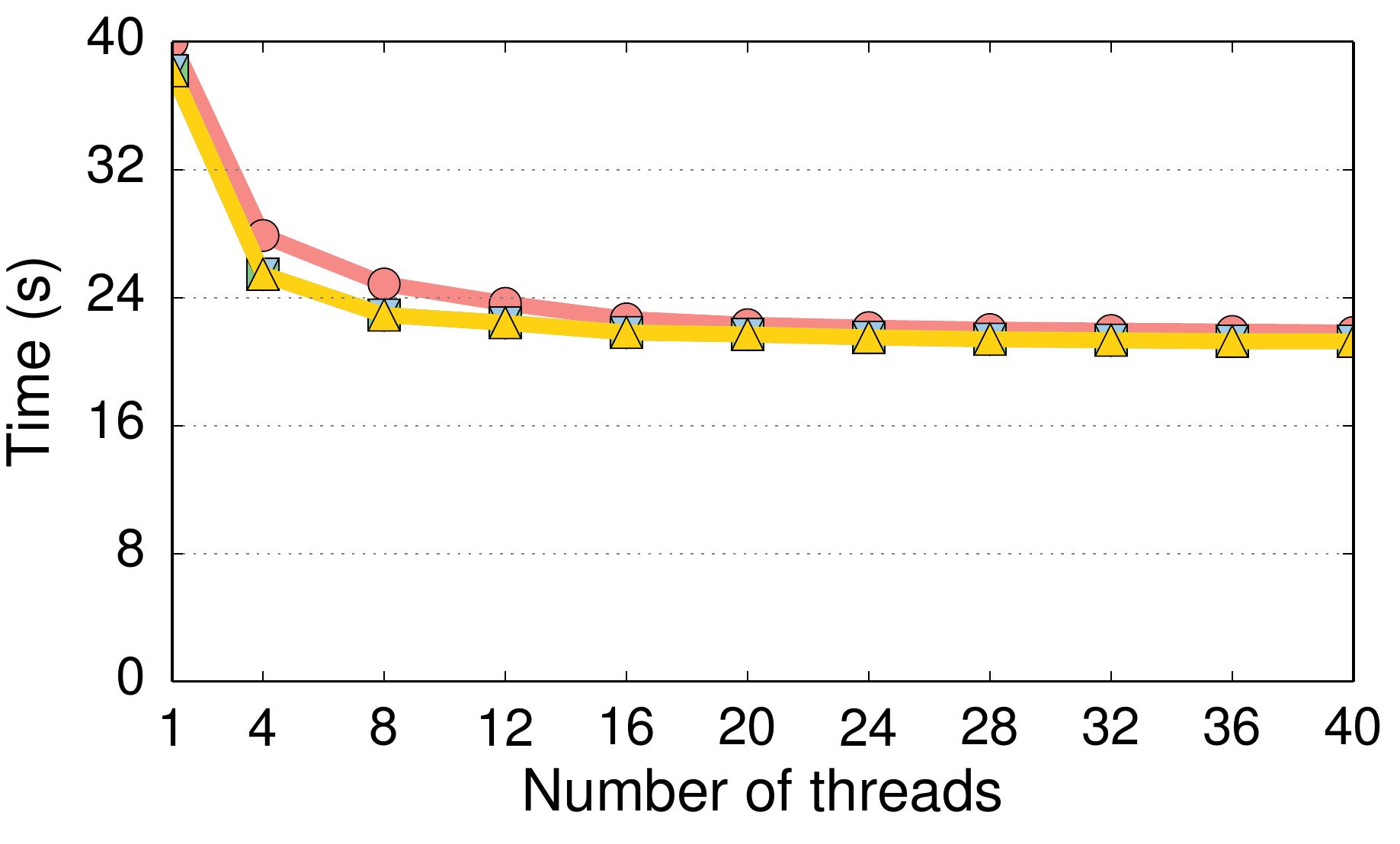}
        \label{figure:exp:tpcc_ckpt_LD}
    }
    \subfloat[Overall time duration.]{
        \includegraphics[width=0.49\columnwidth]
            {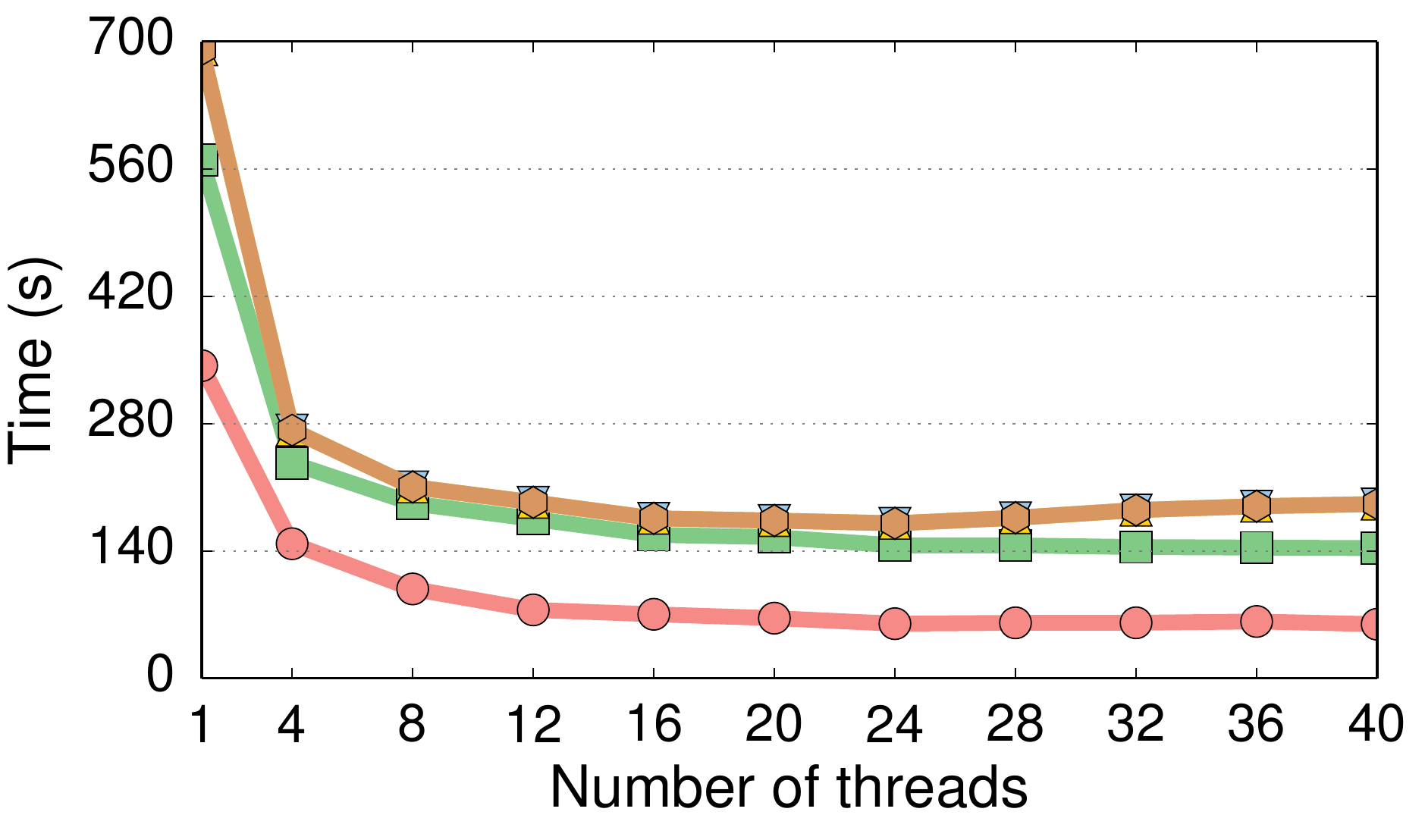}
        \label{figure:exp:tpcc_ckpt_LD_DS_TR}
    }
    \caption{
        Performance of checkpoint recovery.
    }
    \label{figure:exp:checkpoint_recovery}
\end{figure}

We first examine the performance of each scheme's checkpoint recovery stage.
We use the TPC-C benchmark and require the DBMS to recover a 20 GB database state.
\mbox{\cref{figure:exp:tpcc_ckpt_LD}} compares the checkpoint file
reloading time of each recovery scheme.
The result shows that different recovery schemes 
require a similar time duration
for reloading checkpoint files from the underlying storage,
and the reloading speed can easily reach the peak bandwidth of the
two underlying SSDs, which is $\sim$1 GB/s.
However, the results in \cref{figure:exp:tpcc_ckpt_LD_DS_TR}
indicate that PLR's checkpointing scheme requires much less time for 
completing the entire checkpoint recovery phase.
This is because this scheme only restores the database records during checkpoint recovery,
and the reconstruction of all the database indexes is performed during the subsequent log recovery phase.
All the other checkpointing schemes, however, must perform on-line index reconstruction,
as their subsequent log recovery phase needs to use the indexes for tuple retrievals.
LLR's checkpoint recovery scheme also perform slightly faster than the rest ones,
as it can leverage multi-versioning to increase the recovery concurrency.

\subsubsection{Log Recovery}

We now compare each scheme's log recovery stage using the TPC-C benchmark.
The recovery process was triggered by crashing the DBMS after the benchmark
has been executed for 5 minutes.

\begin{figure}[t!]
    \centering
    \fbox{
    \includegraphics[width=0.9\columnwidth]
        {figs-exp/recovery/recovery_legend.pdf}
    }
    \subfloat[Pure log file reloading.]{
        \includegraphics[width=0.49\columnwidth]
            {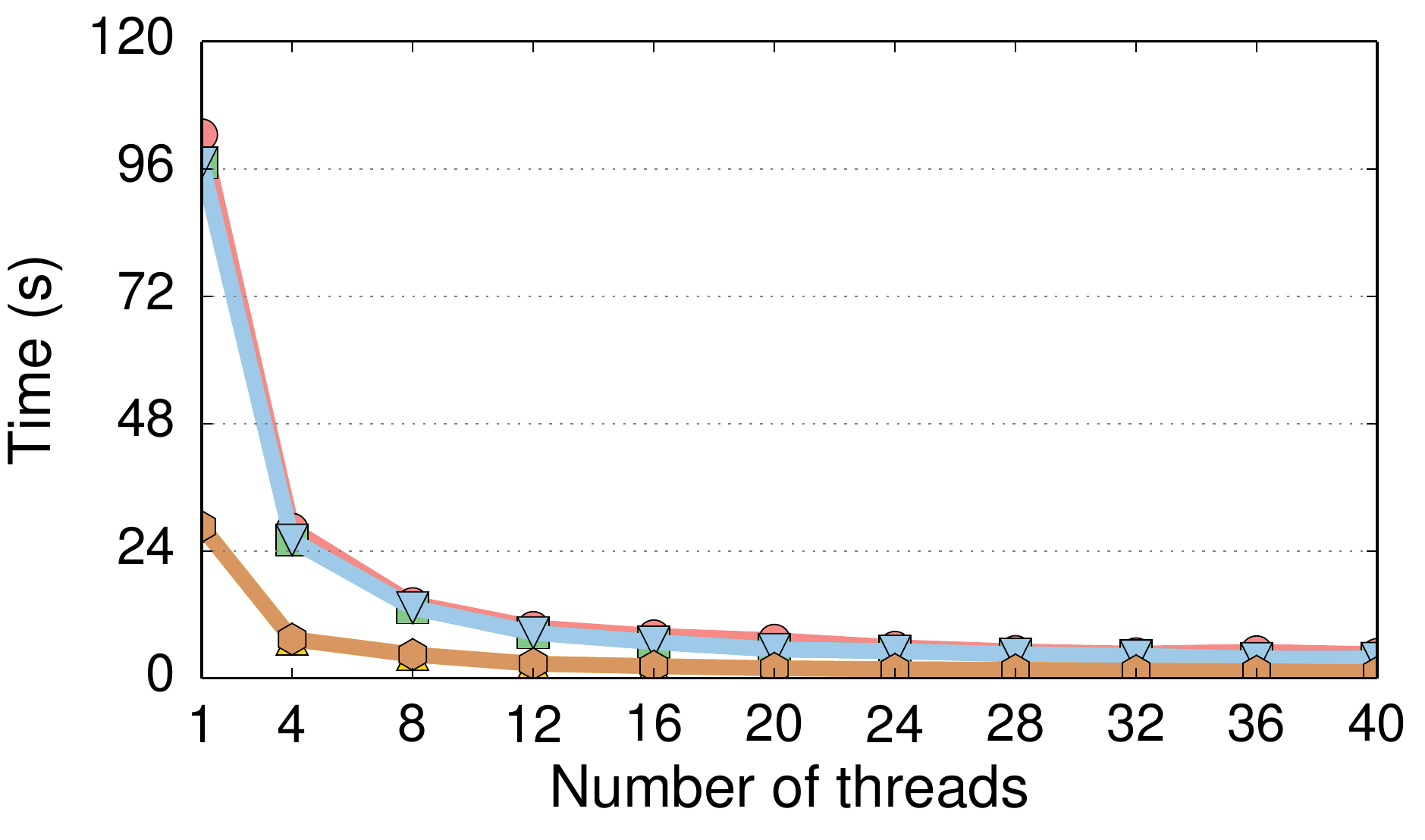}
        \label{figure:exp:tpcc_log_LD}
    }
    \subfloat[Overall time duration.]{
        \includegraphics[width=0.49\columnwidth]
            {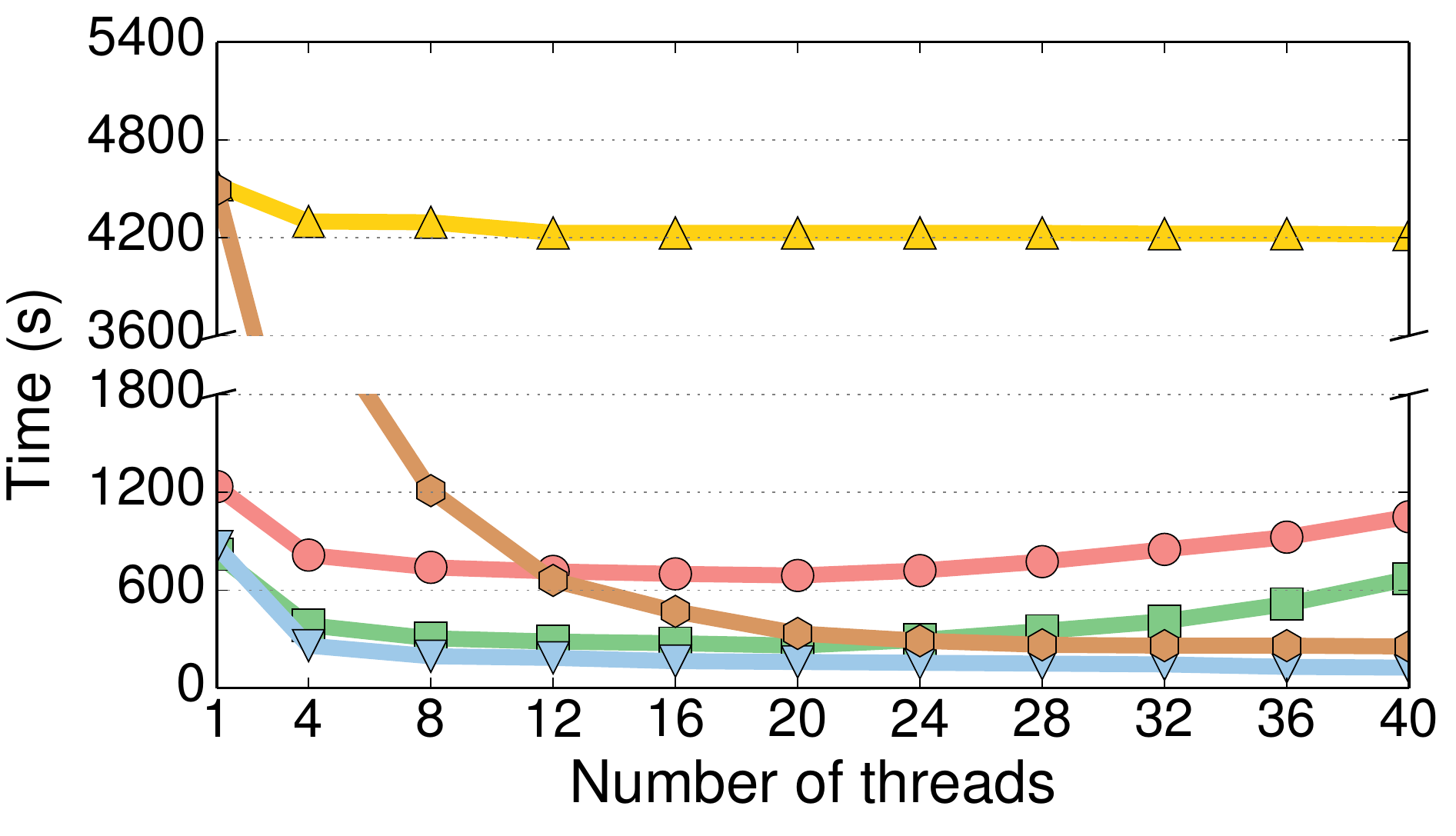}
        \label{figure:exp:tpcc_log_LD_DS_TR_IR}
    }
    \caption{
        Performance of log recovery.
    }
    \label{figure:exp:log_recovery}
\end{figure}

\cref{figure:exp:tpcc_log_LD} shows the recovery time of each log recovery scheme.
Compared to the tuple-level log recovery schemes (i.e., PLR, LLR, and LLR-P), 
the transaction-level log recovery schemes (i.e., CLR and CLR-P) require much less time for
log reloading. This is because transaction-level logging can generate much smaller log files
compared to tuple-level logging, especially when processing write-intensive workloads (like TPC-C).


\cref{figure:exp:tpcc_log_LD_DS_TR_IR} also demonstrates the significant performance improvement of CLR-P over CLR.
As CLR utilizes only a single thread for log replay,
CLR took over 4,200 seconds (70 minutes) to complete the log recovery.
In contrast, by utilizing multiple threads for recovery,
our proposed CLR-P was able to outperform CLR by a factor of 18.
Observe that the performance of CLR-P improves significantly with the number of recovery threads.
As CLR-P already schedules the transaction re-execution order (using both static and dynamic analyses), 
CLR-P does not require latching during recovery and therefore is not hampered
by the latch synchronization overhead inherent in CLR.

Observe that for both PLR and LLR,
their recovery times improve with the number of recovery threads up to 20 threads
and beyond that point, their recovery times increase with the number of recovery threads.  
This is because the recovery threads in both PLR and LLR (which follow SiloR's design) require latches on tuples to be modified for recovery correctness,
and the synchronization overhead of using latches start to degrade the overall performance beyond 20 recovery threads.

\begin{figure}[t!]
    \centering
    \fbox{
    \includegraphics[width=0.9\columnwidth]
        {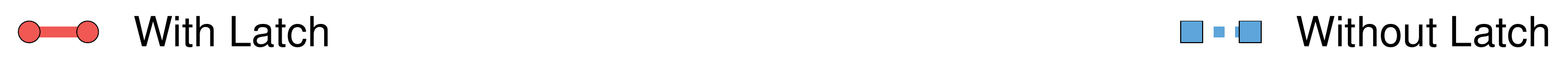}
    }
    \subfloat[PLR.]{
        \includegraphics[width=0.49\columnwidth]
            {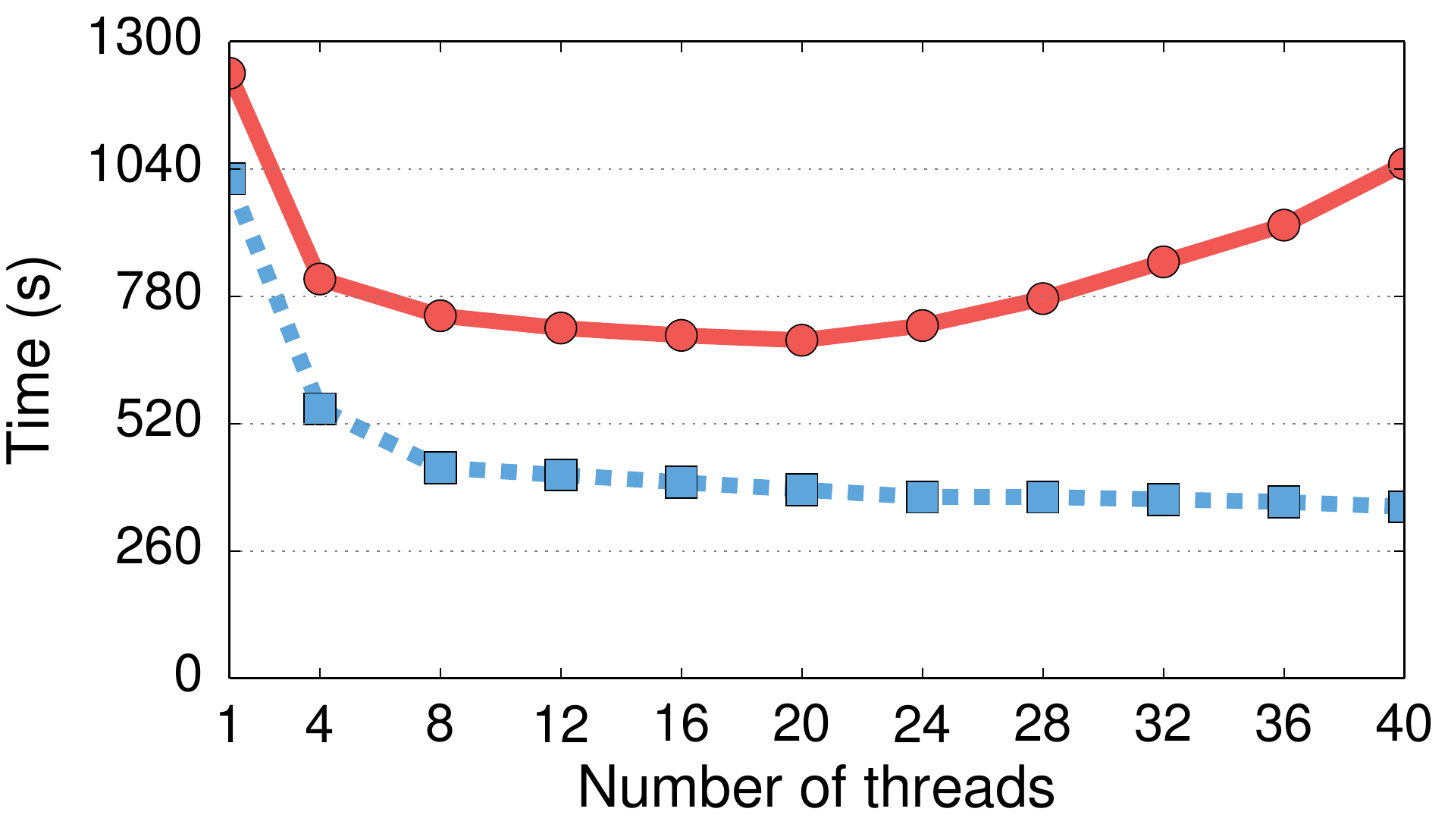}
        \label{figure:exp:tpcc_physical_compare}
    }
    \subfloat[LLR.]{
        \includegraphics[width=0.49\columnwidth]
            {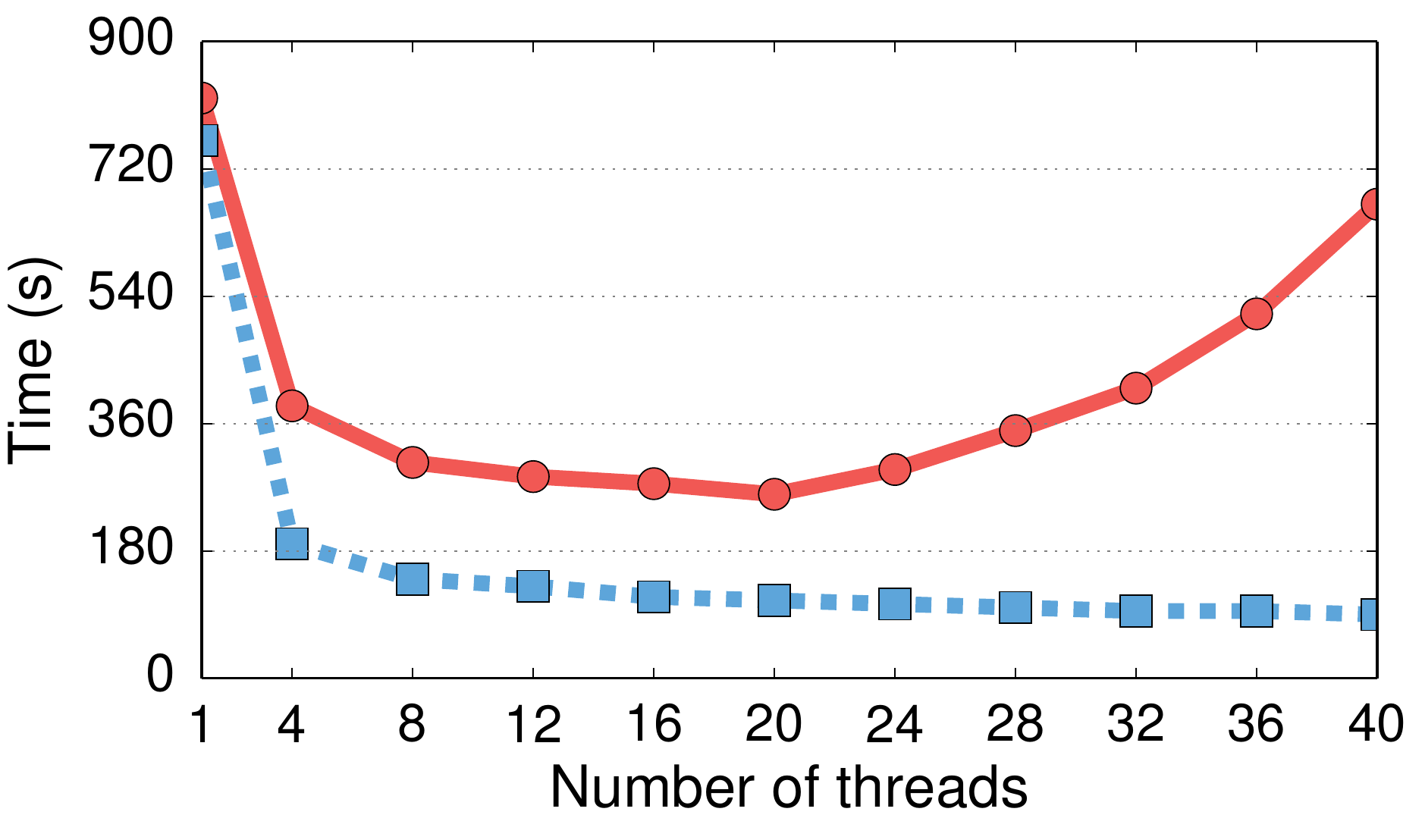}
        \label{figure:exp:tpcc_logical_compare}
    }
    \caption{
        Latching Bottleneck in tuple-level log recovery schemes.
    }
    \label{figure:exp:tpcc_recovery_compare}
\end{figure}

To try to quantify the latching overhead incurred by PLR and LLR, 
we removed the latch acquisition operations in both of these recovery
schemes and then measured their recovery performance.
Of course, without the use of latches,
both PLR and LLR could produce inconsistent database states after recovery;
however, the attained performance measurements would essentially 
indicate the peak performance achievable by PLR and LLR.
As shown in \cref{figure:exp:tpcc_recovery_compare}, 
with the latch acquisition disabled, 
the recovery times of both PLR and LLR drop significantly with 
the increase in the number of recovery threads.
Observe that the time reduction after 12 threads is not quite significant. 
This is because
(1) the scalability of the log reloading phase is bounded by the maximum 
read throughput of the underlying SSD storage; 
and (2) the scalability of the log replay phase is also constrained by 
the performance of the concurrent database indexes.
With 20 recovery threads, the recovery times of PLR and LLR were 
reduced to the minimum at around 750 and 270 seconds respectively. 
However, scaling these two schemes towards 40 threads significantly increases the
recovery time to over 1000 and 700 seconds, respectively. 
These results show the inefficiency of the state-of-the-art tuple-level log recovery schemes.

\subsubsection{Overall Performance}

This section evaluates the overall performance of the recovery schemes
using 40 recovery threads.
As before, the recovery schemes were triggered after 5 minutes of transaction processing.

\begin{figure}[t!]
    \centering
    \fbox{
    \includegraphics[width=0.9\columnwidth]
        {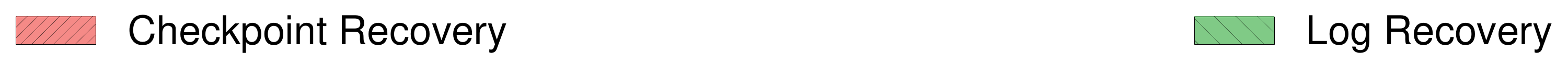}
    }
    \subfloat[TPC-C.]{
        \includegraphics[width=0.49\columnwidth]
            {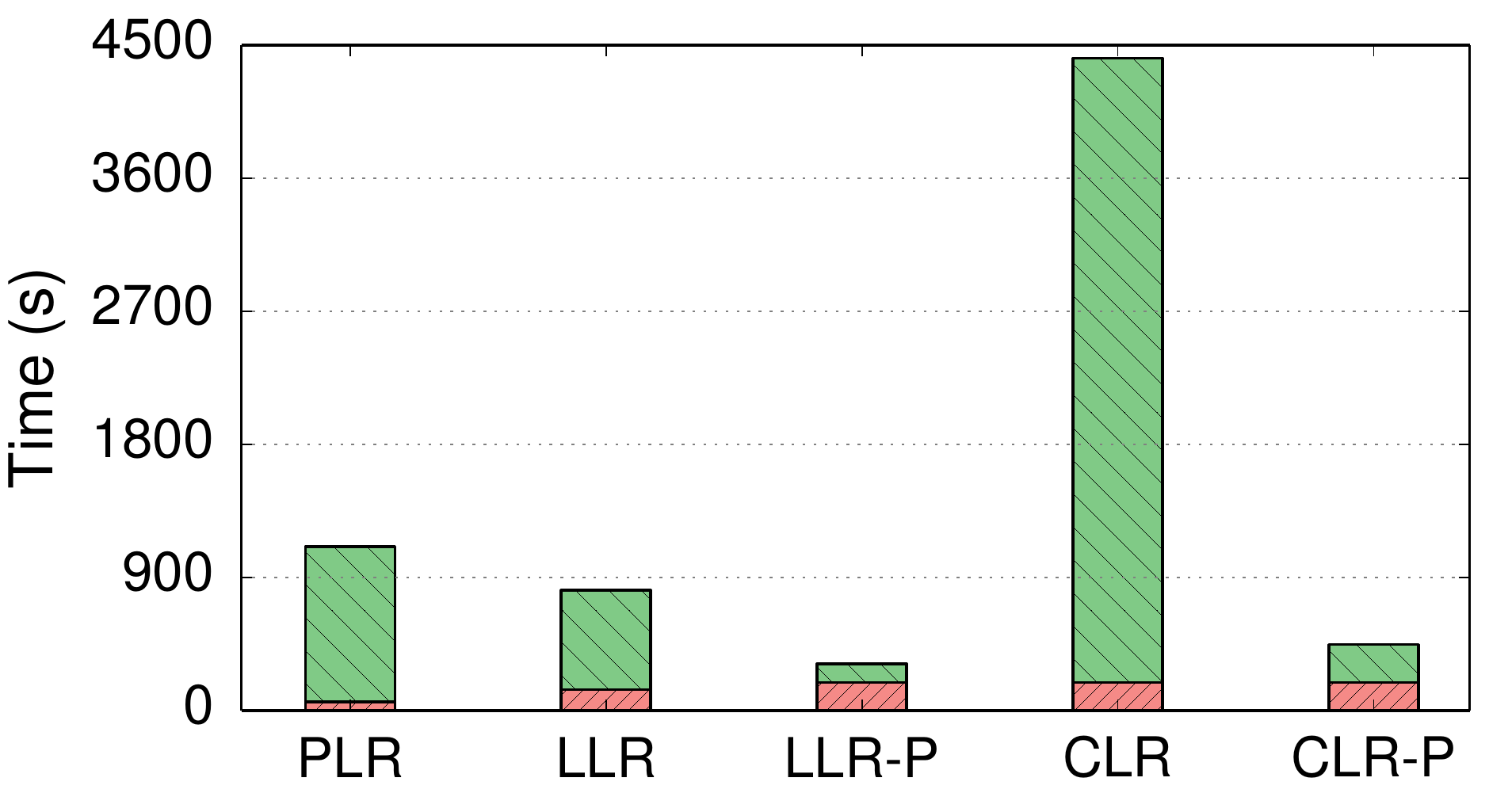}
        \label{figure:exp:tpcc_overall_breakdown}
    }
    \subfloat[Smallbank.]{
        \includegraphics[width=0.49\columnwidth]
            {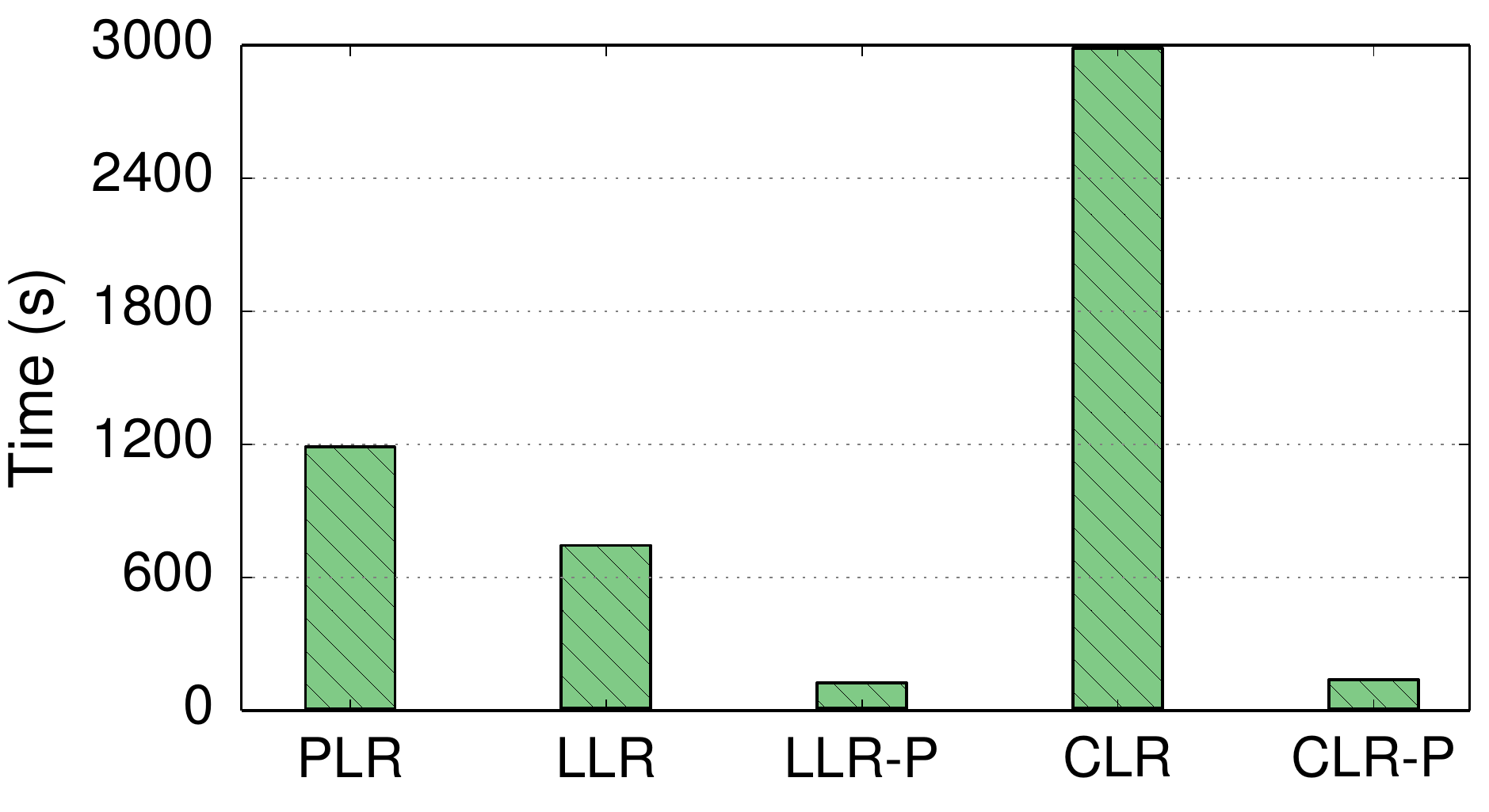}
        \label{figure:exp:smallbank_overall_breakdown}
    }
    \caption{
        Overall performance of database recovery.
    }
    \label{figure:exp:overall_breakdown}
\end{figure}

As shown in \cref{figure:exp:overall_breakdown},
CLR performed the worst in both benchmarks as CLR
cannot leverage multi-threading for reducing log recovery time.
Our proposed scheme, LLR-P, achieved the best performance.
This is due to two main reasons. 
First, unlike CLR, LLR-P is able to exploit multiple recovery threads for efficient recovery.
Second, 
LLR-P schedules the transaction re-execution order beforehand and it does not require any latching thereby avoiding the synchronization overhead
that is incurred by both PLR and LLR schemes.
We note that CLR-P consumes more time than LLR-P for recovering the database. This is because
CLR-P has to re-execute all the operations (including both read and write) in a transaction,
whereas LLR-P only reinstalls modifications recorded in the log files.
For all the compared schemes, the checkpoint recovery time is almost negligible, 
as this phase is easily parallelized.

\subsubsection{Ad-Hoc Transactions}

\begin{figure}[t!]
    \centering
    \fbox{
    \includegraphics[width=0.9\columnwidth]
        {figs-exp/recovery/recovery_overall_legend.pdf}
    }
    \subfloat[TPC-C.]{
        \includegraphics[width=0.49\columnwidth]
            {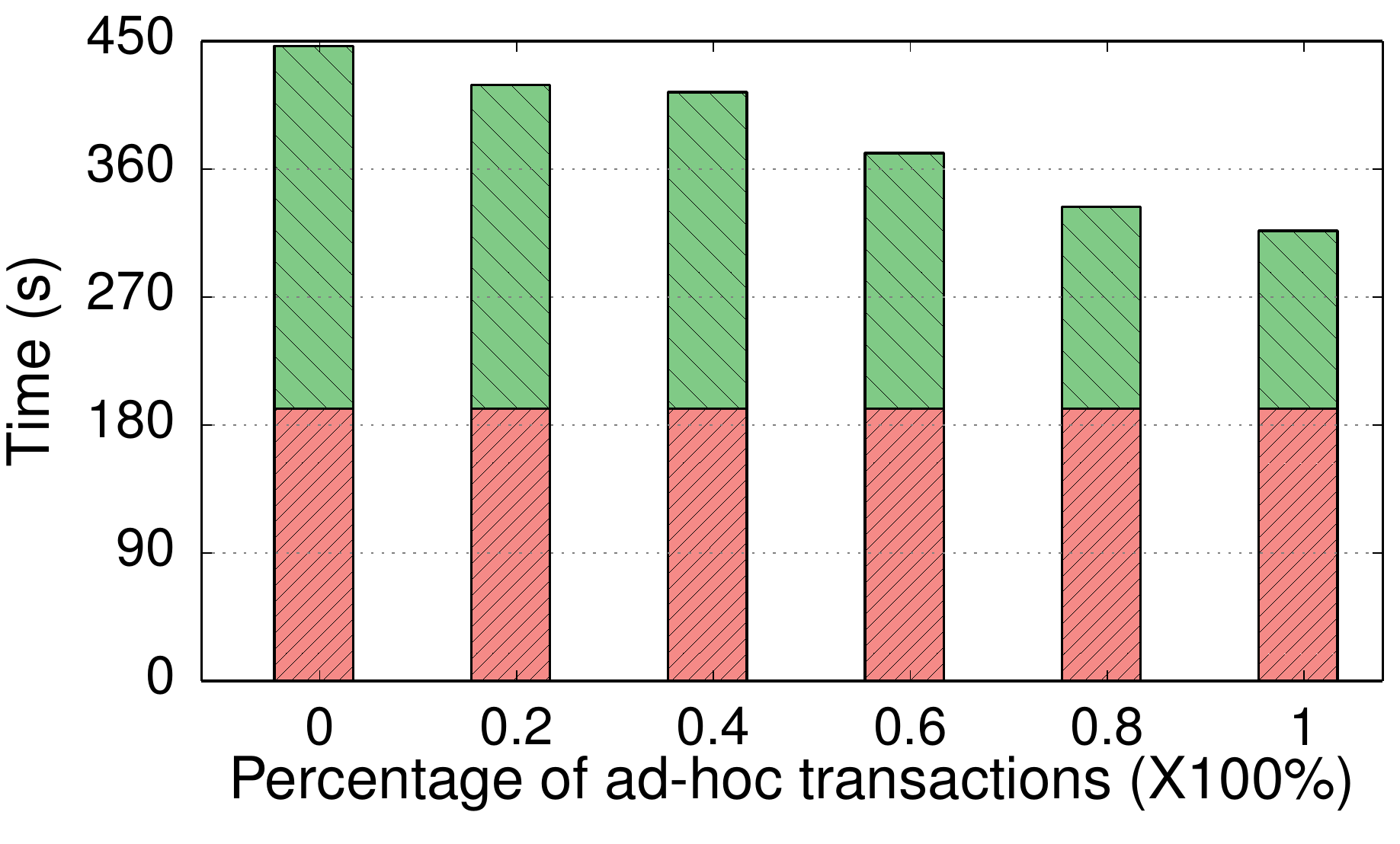}
        \label{figure:exp:tpcc_recovery_adhoc}
    }
    \subfloat[Smallbank.]{
        \includegraphics[width=0.49\columnwidth]
            {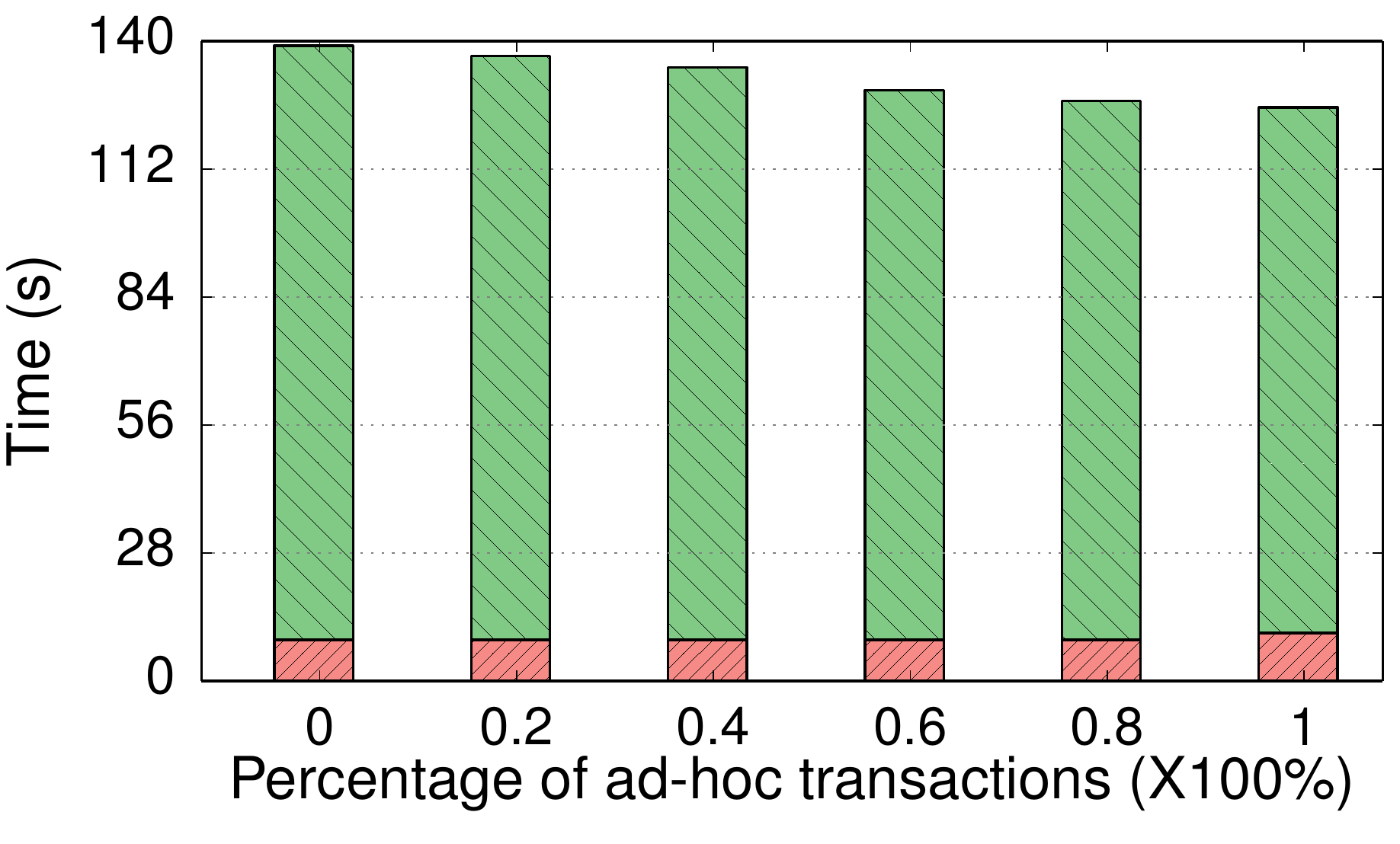}
        \label{figure:exp:smallbank_recovery_adhoc}
    }
    \caption{
        Database recovery with ad-hoc transactions.
    }
    \label{figure:exp:recovery_adhoc}
\end{figure}

We further measure how the presence of ad-hoc transactions 
influence \system's performance in database recovery.
We use the same configurations as the previous experiments, and mix the workload with certain
percentage of ad-hoc transactions.
\cref{figure:exp:recovery_adhoc} shows the results.
By varying the percentage of ad-hoc transactions from 0\% to 100\%, 
the recovery time of \system drops smoothly.
When the percentage of ad-hoc transactions is increased to 100\%, this
result essentially show the performance of LLR-P. As recovering command logs requires the DBMS
to perform all the read operations in the stored procedure, it takes more time compared
to pure logical log recovery.
This results confirmed the efficiency of \system's support of ad-hoc transactions.

The experiment results reported in this section confirmed that 
\system requires a much lower recovery time for restoring lost database states 
compared with the state-of-the-art recovery schemes, even in the existence
of ad-hoc transactions.

\subsection{Performance Analysis}

In this section, we analyze the effectiveness of each of the proposed mechanisms in \system using the TPC-C benchmark.
In particular, we measure the recovery performance
achieved by \system's static analysis and dynamic analysis,
and then investigate the potential performance bottlenecks in \system.

The results reported in this section are based on running the benchmark
for a duration of five minutes and then triggering a database crash to start
the recovery process.
As both static and dynamic analyses are designed for log recovery,
we omit checkpoint recovery in this section's experiments.

\subsubsection{Static Analysis}


As the static analysis in \system relies on decomposing stored procedures into slices to enable execution parallelism, 
we compare the effectiveness of \system's decomposition technique
against a baseline technique that is adapted from the well-known transaction chopping technique~\cite{shasha1995transaction}.
A qualitative comparison of these two techniques is given in 
\cref{relatedwork}.

\begin{figure}[t!]
	\centering
    \fbox{
    \includegraphics[width=0.9\columnwidth]
        {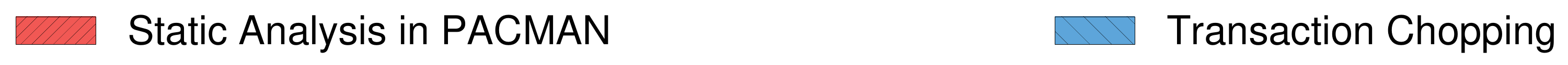}
    }
	\includegraphics[width=0.7\columnwidth,clip]
		{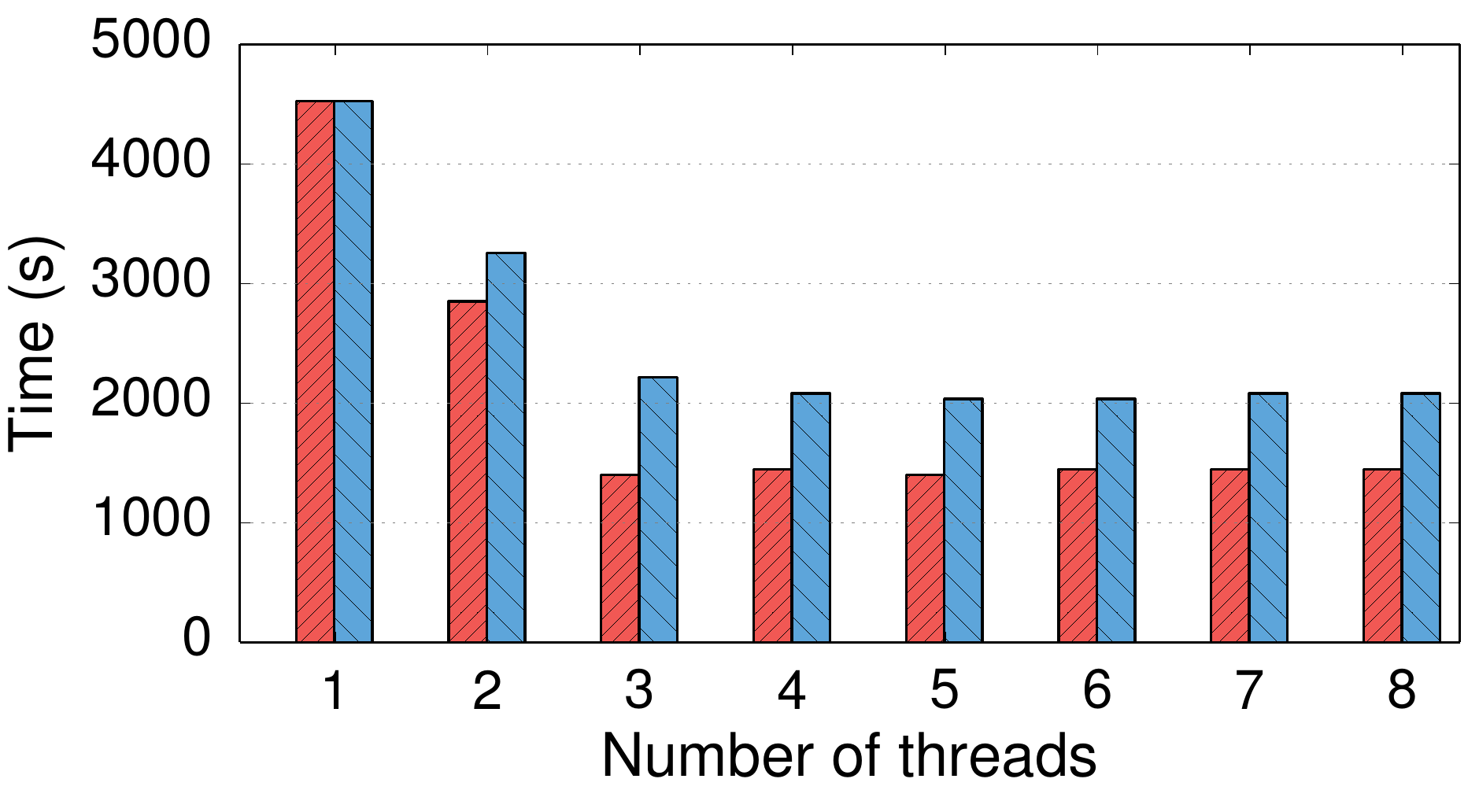}
	\caption{Effectiveness of static analysis.}
	\label{figure:exp:static_analysis}
\end{figure}

\cref{figure:exp:static_analysis} compares the log recovery performance achieved
by \system's static analysis and the transaction chopping-based scheme.
For this experiment, the dynamic analysis phase was disabled 
to focus on the comparison between the two competing static analysis techniques.
The results show that,
as the number of threads increases from 1 to 3, 
the recovery time achieved by \system's static analysis decreases from 4500 seconds to $\sim$2000 seconds. 
But beyond this point, the recovery time stops decreasing and there is no further 
performance gain brought from the increased thread count. 
This is because \system's static analysis 
extracts only coarse-grained parallelism for log recovery, 
and dynamic analysis needs to be incorporated to 
fully exploit the multi-thread execution. 
The same figure also shows the recovery time required by 
transaction chopping
is always longer than that required by \system's static analysis. 
This is because the decomposition obtained from \system is finer-grained than
that from transaction chopping.
\eat{
We also note that \system's full performance 
is further facilitated by the dynamic analysis applied at runtime. 
In contrast, the adapted transaction chopping approach is unable to incorporate runtime information to further lower the recovery time.
}

\eat{
\cref{figure:exp:static_analysis} compares the log recovery performance achieved
by \system's pure static analysis and a transaction chopping-based scheme, 
which is adapted to database recovery by performing procedure analysis 
and enforcing transaction ordering (see \cref{sec:chopping}).
To the best of our knowledge, transaction chopping is the only static analysis-based solution 
that is applicable for database recovery.
The results show that,
as the number of threads increases from 1 to 3, 
the recovery time achieved by \system's static analysis decreases from 4500 seconds to $\sim$2000 seconds. 
But beyond this point, the recovery time stops decreasing and there is no further 
performance gain brought from the increased thread count. 
This is because \system's static analysis 
extracts only coarse-grained parallelism for log recovery, 
and dynamic analysis needs to be incorporated to fully utilize 
the computing power of multi-thread execution. 
The same figure also shows the recovery time required by transaction chopping
is always longer than that required by \system's static analysis. 
This is because \system can delay the determination of re-execution order to
the database recovery time, and hence the extracted parallelism at static time 
is finer-grained. 
We also note that \system's full performance 
is further facilitated by the dynamic analysis applied at runtime. 
In contrast, the adapted transaction chopping approach is unable to incorporate runtime information 
to further lower the recovery time.
}


\begin{figure}[t!]
  \centering
    \fbox{
    \includegraphics[width=0.9\columnwidth]
        {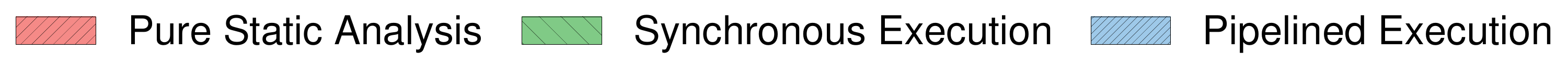}
    }
  \includegraphics[width=0.7\columnwidth,clip]
    {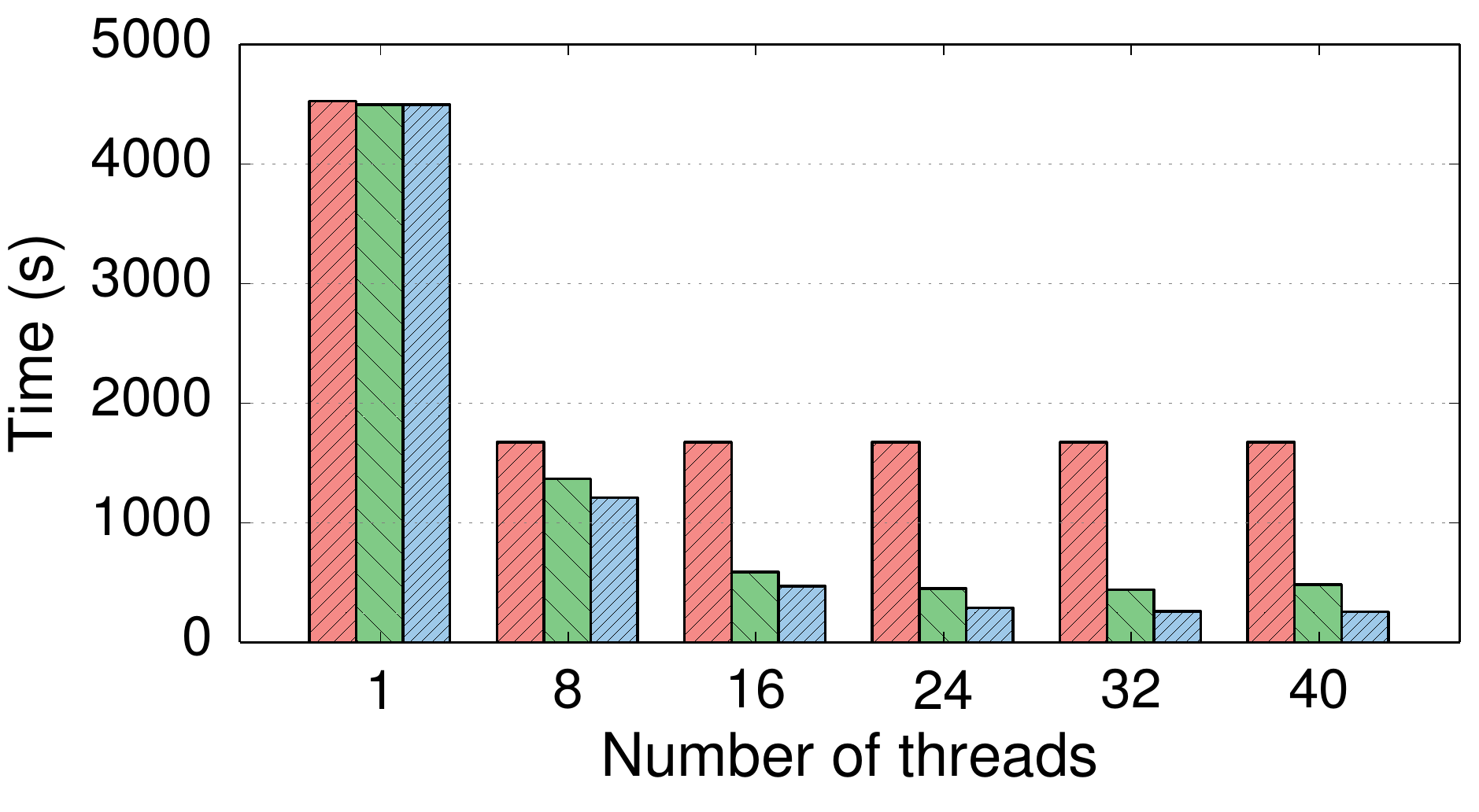}
    \caption{Effectiveness of dynamic analysis.}
  \label{figure:exp:dynamic_analysis}
\end{figure}

\subsubsection{Dynamic Analysis}


This section examines the effectiveness of the dynamic analysis in \system. 
We analyze the benefits of intra- and inter-batch parallelism  by comparing three techniques:
(1) using only static analysis techniques (without applying any techniques from dynamic analysis),
(2) using techniques from both static analysis and intra-batch parallelism techniques (i.e., synchronous execution),
and
(3) using all the techniques from static and dynamic analyses (i.e.,  pipelined execution).
\cref{figure:exp:dynamic_analysis} shows that, 
by using synchronous execution,
\system yields over 4 times lower recovery time 
compared to that achieved by pure static analysis with 40 threads enabled. 
The performance is further improved by exploiting inter-batch parallelism. 
Specifically, with pipelined execution,
the recovery time of \system drops to less than 300 seconds when utilizing 40 threads. 
This result confirms that both the intra- and inter-batch parallelism extracted in 
\system can help improve the system scalability and hence reduce recovery time.

\subsubsection{Time Breakdown}

\begin{figure}[t!]
	\centering
    \fbox{
    \includegraphics[width=0.9\columnwidth]
        {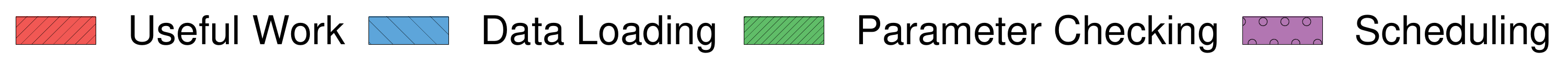}
    }
	\includegraphics[width=0.7\columnwidth,clip]
		{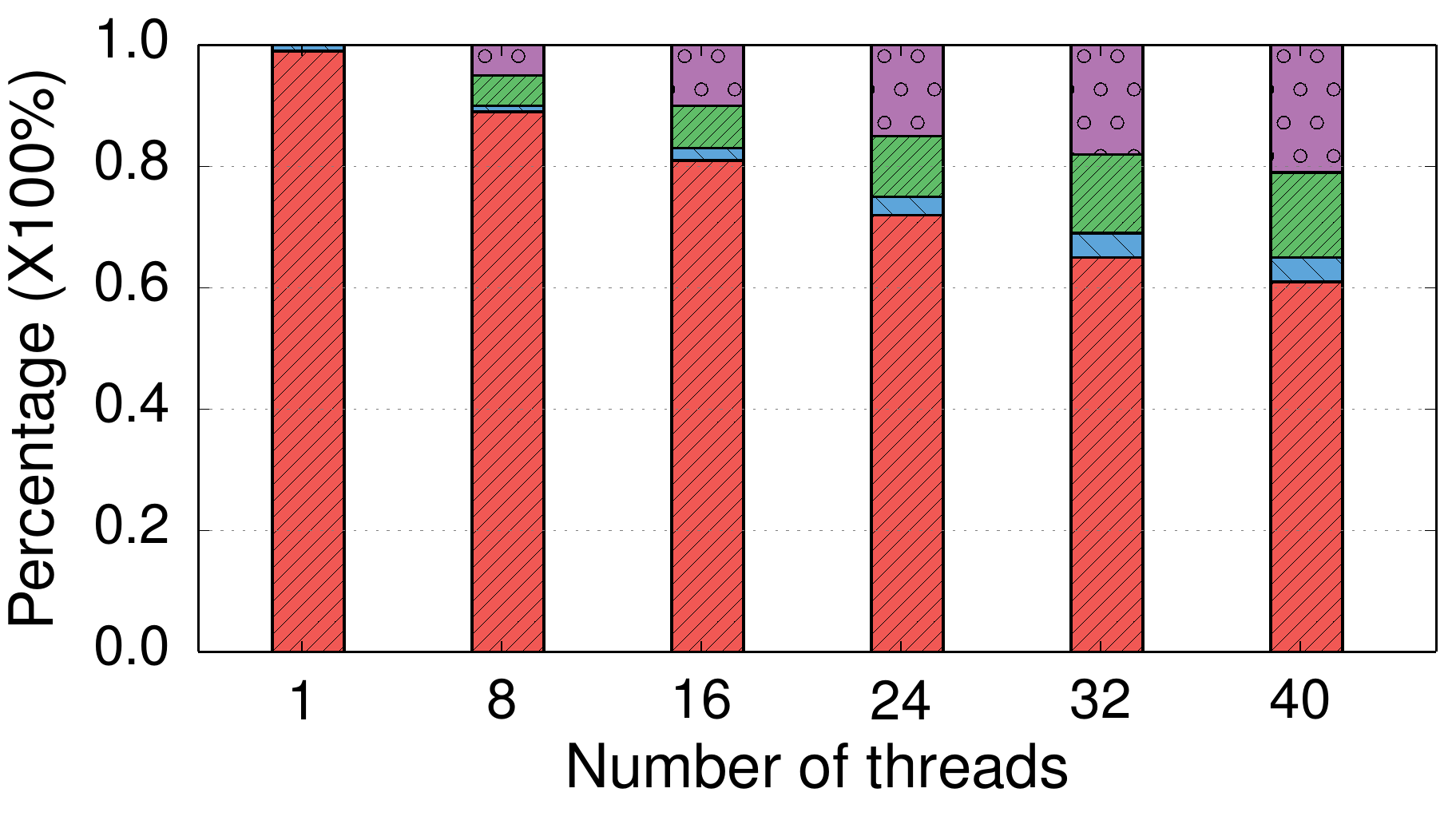}
		\caption{Log recovery time breakdown.}
	\label{figure:exp:time_breakdown}
\end{figure}

Having understood how each of the proposed mechanisms contributes to the system performance, 
we further investigate the performance bottleneck of \system. 
The bottleneck can potentially come from three sources. 
First, the DBMS needs to load the log files from the underlying storage and deserialize
the logs to the main-memory data structures.
Second, the dynamic analysis in \system requires that the parameter values in each log batch be analyzed for deriving intra-batch parallelism, possibly blocking the subsequent tasks.
Third, the scheduling of multiple threads requires each thread to 
access a centralized data structure, potentially resulting in intensive data races.
We break down the recovery time of \system and show the result in \cref{figure:exp:time_breakdown}.
By scaling \system to 40 threads, thread scheduling becomes the major bottleneck, 
occupying around 30\% of the total recovery time. 
In contrast, log data loading and dynamic analysis are very lightweight, 
and these two processes do not lead to high overhead. 
Observing the performance bottleneck in \system, 
we argue that employing a better scheduling mechanism can help 
further optimize the performance of database recovery. 

\section{Related Work}
\label{relatedwork}


Main-memory DBMSs have 
been well studied by the research community for over two decades~\cite{dewitt1984implementation,grund2010hyrise,kallman2008h,kemper2011hyper,tu2013speedy,wu2016scalable}. 
Database recovery for such DBMSs use a combination of checkpointing and logging mechanisms. 
While there have been many recent approaches on improving the performance of checkpointing~\cite{cao2011fast,liedes2006siren,Ren2016low,salem1989checkpointing},
several previous works~\cite{malviya2014rethinking,zheng2014fast},
as well as our study in this paper,
have shown that log recovery is the major bottleneck for database recovery.

Log-based recovery techniques face a performance trade-off 
between transaction processing and failure recovery. 
While tuple-level logging~\cite{mohan1992aries} offers faster recovery
than transaction-level logging~\cite{lomet2011implementing,malviya2014rethinking},
the latter incurs lower overhead during normal transaction processing.
Existing works largely focused on optimizing tuple-level logging mechanisms 
with techniques such as log compression~\cite{dewitt1984implementation,li1993post} 
and using hardware support~\cite{johnson2010aether,ongaro2011fast,wang2014scalable,zheng2014fast}. 
A recent work by Yao et al.~\cite{yao2016adaptive} investigated the recovery costs 
between transaction-level and tuple-level logging for distributed in-memory DBMSs. 
As a significant departure from existing works, 
our work on \system focuses on achieving high performance in both the logging and recovery 
processes for the transaction-level logging approach. 

The idea behind \system is inspired a series of recent works
that leverage transaction analysis for advanced performance. 
For example,
Doppel~\cite{narula2014phase} execute commutative operations in parallel for higher transaction-processing throughput.
Yan et al.~\cite{yan2016leveraging} extracted data dependencies within transactions
to improve transaction
processing performance under high-contention workloads.
Wu et al.~\cite{wu2016transaction} analyzed dependencies within each transaction to
scale conventional optimistic concurrency control on multicores.
A well-known technique in this area is transaction chopping~\cite{shasha1995transaction},
which tries to increase the concurrency for a given workload of transactions.
By analyzing the conflicting operations among the transactions in the workload,
each transaction is decomposed into a set of smaller sub-transactions
such that any strict two-phase locking execution of the collection of sub-transactions
is a serializable execution (w.r.t. to the original workload of non-decomposed transactions).  
Several recent works have applied transaction chopping to optimize the processing of 
distributed transactions~\cite{mu2014extracting,zhang2013transaction}.

Similar to the use of conflicting operations for decomposing transactions in transaction chopping, 
the static analysis in \mbox{\system} uses flow dependencies to decompose stored procedures into slices.
However, a key difference between these techniques
is that they are developed for different objectives that have different constraints.
The goal of transaction decompositions in \mbox{\system} is to parallelize
the replay of committed transactions during database recovery,
and thus the execution order of the decomposed transaction pieces is chosen to maximize execution parallelism
while respecting the ordering constraints from the flow dependencies among the transaction operations
and that from the transactions in the recovery log. 
In contrast, transaction chopping is designed to maximize concurrency during normal transaction executions
and its decomposition needs to satisfy a different and stronger property that any strict 2PL execution of the decomposed
sub-transactions is serializable. Consequently, the granularity of the decompositions from transaction chopping are coarser
than those from \mbox{\system}.

The techniques used in our dynamic analysis share some similarities with concurrency control techniques
in that they both aim to find opportunities for inter-transaction parallelism.
However, a key difference between these techniques is the context in which they operate. 
In the context of \mbox{\system} for database recovery, 
the set of committed transactions to be replayed are known before the start of recovery
and the input parameter values for the transactions are also known from the recovery log.
Consequently, \mbox{\system} is able to exploit more information to maximize execution parallelism.
In contrast, conventional concurrency control techniques are applied in a more dynamic context 
where the order of incoming transaction operations is not known apriori and thus the parallelism opportunities
are more limited.


\section{Conclusion}
\label{conclusion}
We have developed \system, a database recovery mechanism that 
achieves speedy failure recovery without introducing any costly overhead to the transaction processing.
By leveraging a combination of static and dynamic analyses, 
\system exploits fine-grained parallelism for replaying logs generated by
coarse-grained transaction-level logging.
By performing extensive performance studies on
a 40-core machine, we confirmed that 
\system can significantly reduce the 
database recovery time
compared to the state-of-the-art recovery schemes.


\newpage

\bibliographystyle{abbrv}
{\small \bibliography{sigproc}}  


\appendix

\section{Implementation}
\label{implementation}

In this section, we describe the implementation details
of the logging-and-recovery framework adopted in \database.
Our implementation faithfully follows that of SiloR~\cite{zheng2014fast}, 
a main-memory DBMS that is optimized for fast durability. 
We discuss some possible optimization techniques
at the end of this section.

\subsection{Logging}
The DBMS spawns a collection of worker threads for processing
transactions and a collection of logger threads for
persisting logs.
Worker threads are divided into multiple sub-groups, each of 
which is mapped to a single logger thread.

To minimize the logging overhead brought by frequent disk accesses,
the DBMS adopts group commit scheme and persists 
logs in units of epochs. 
This requires each logger thread to 
pack together all its transaction logs generated in a certain epoch
before flushing them into the secondary storage.
To limit the file size and facilitate log recovery,
a logger thread truncates its corresponding log sequence
into a series of finite-size log batches, and each batch contains
log entries generated in multiple epochs.
The DBMS stores different log batches in different log files,
and this mechanism simplifies the process of locating log entries
during log recovery.

Each logger thread in the DBMS works independently, and this
requires us to create a new thread, called \textit{pepoch} thread,
to continuously detect the slowest progress of these logger threads.
If all the loggers have finished persisting epoch $i$, then
the pepoch thread writes the number $i$ into a file named 
\texttt{pepoch.log} and notifies all the workers that query results 
generated for any transaction before epoch $i + 1$ can
be returned to the clients.

and the batch size to 100 epochs.

\subsection{Recovery}
The DBMS starts log recovery by first reading the latest 
persisted epoch ID maintained in the file \texttt{pepoch.log}.
After obtaining the epoch ID, the DBMS reloads 
the corresponding log files and replays the persisted log entries. 
For tuple-level logging mechanisms,
including physical logging and logical logging, the DBMS replays
the log files in the reverse order than they were written.
This mechanism minimizes the overhead brought by data copy.
However, for transaction-level logging mechanism, or command logging,
the DBMS has to replay transaction logs following the transaction
commitment order, as described in this paper.

\subsection{Possible Optimizations}
Existing works have proposed several mechanisms for optimizing
the performance of logging-and-recovery mechanism in DBMSs.
However, these optimizations may not be suitable for main-memory DBMSs.

A widely used optimization mechanism in disk-based DBMSs is 
log compression~\cite{dewitt1984implementation,li1993post},
which aims at minimizing the log size that is dumped to the disk.
We did not adopt this mechanism, as SiloR's experiments
have shown that compression can degrade the logging performance in
main-memory DBMSs~\cite{zheng2014fast}.
Some DBMSs adopt delta logging~\cite{schwalb2015efficient} or 
differential logging~\cite{lee2001differential} to persist only 
the updated columns of the tuples for a transaction.
While reducing the log size, these mechanisms are specifically designed for
multi-version DBMSs. We did not adopt these optimization schemes,
as our goal is to provide a generalized logging
mechanism for both single-version and multi-version main-memory DBMSs.
Kim et al.~\cite{kim2016ermia} implemented a
latch-free scheme to achieve scalable centralized logging in 
a main-memory DBMS called Ermia.
Their mechanism is designed for DBMSs that execute transactions
at snapshot isolation level.
We keep using SiloR's design as \database
provides full serializability for transaction processing.
Hekaton~\cite{diaconu2013hekaton}'s logging implementation is
very similar to ours, and it also avoids write-ahead logging
and adopts group commit to minimize overhead from disk accesses.
We have already included its optimization schemes in our implementation.

\section{Algorithms}
\label{sec:algorithms}

This section presents the algorithms for constructing two statically extracted graphs:
local dependency graph (shown in \cref{alg:localgraph}) and global dependency graph (shown in \cref{alg:globalgraph}).

\begin{algorithm}[h!]
\SetKwInOut{Input}{Input}
\SetKwInOut{Output}{Output}
\SetAlgoVlined
\Input{a sequence of operations $O=\{o_1, o_2, \dots, o_n\}$ in a stored procedure $p$}
\Output{a local dependency graph $g$ containing a set of slices $S=\{s_1, s_2, \dots, s_m\}$}
\vspace{1.5mm}
\textbf{Initialization:}\\
$S=\{\{o_i\}$ $|$ $o_i$ is an operation in $O\}$\;
\vspace{1.5mm}
\textbf{Merge slices:}\\
\While{exists $o_p$ and $o_q$ respectively from $s_i$ and $s_j$ that are data-dependent}{
    merge $s_i$ and $s_j$ into a new slice $s_k$\;
}
\vspace{1.5mm}
\textbf{Build graph:}\\
\ForEach{slice pair $\langle s_i, s_j\rangle$ in $S$}{
    \If{exists $o_p$ and $o_q$ respectively from $s_i$ and $s_j$ where $o_q$ is flow-dependent on $o_p$}{
        add a dependency edge from $s_i$ to $s_j$\;
    }
}
\vspace{1.5mm}
\textbf{Break cycles:}\\
\ForEach{slice pair $\langle s_i, s_j\rangle$ in $S$}{
    \If{$s_i$ and $s_j$ are mutually (indirectly) dependent}{
        merge $s_i$ and $s_j$ into a new slice $s_k$\;
    }
}
\caption{Build local dependency graph.}
\label{alg:localgraph}
\end{algorithm}

\begin{algorithm}[h!]
\SetKwInOut{Input}{Input}
\SetKwInOut{Output}{Output}
\SetAlgoVlined
\Input{local dependency graphs $G=\{g_1, g_2, \dots, g_n\}$ from each stored procedure}
\Output{a global dependency graph $\mathcal{G}$ containing a set of blocks $B=\{b_1, b_2, \dots, b_m\}$}
\vspace{1.5mm}
\textbf{Initialization:}\\
$B=\{\{s_i\}$ $|$ $s_i$ is a slice of a graph $g_i$ in $G\}$\;
\vspace{1.5mm}
\textbf{Merge blocks:}\\
\While{exists $s_p$ and $s_q$ respectively from $b_i$ and $b_j$ that are data-dependent}{
    merge $b_i$ and $b_j$ into a new block $b_k$\;
}
\vspace{1.5mm}
\textbf{Build graph:}\\
\ForEach{block pair $\langle b_i, b_j\rangle$ in $B$}{
    \If{exists $s_p$ and $s_q$ respectively from $b_i$ and $b_j$ where $s_q$ is dependent on $s_p$}{
        add a dependency edge from $b_i$ to $b_j$\;
    }
}
\vspace{1.5mm}
\textbf{Break cycles:}\\
\ForEach{block pair $\langle b_i, b_j\rangle$ in $B$}{
    \If{$b_i$ and $b_j$ are mutually (indirectly) dependent}{
        merge $b_i$ and $b_j$ into a new block $b_k$\;
    }
}
\vspace{1.5mm}
\textbf{Merge slices:}\\
\ForEach{block $b$ in $B$}{
    merge slices originated from the same stored procedure\;
}
\caption{Build global dependency graph.}
\label{alg:globalgraph}
\end{algorithm}


\section{TPC-C}
\label{sec:tpcc}

\begin{figure}[ht!]
  \centering
  \includegraphics[width=0.7\columnwidth]
    {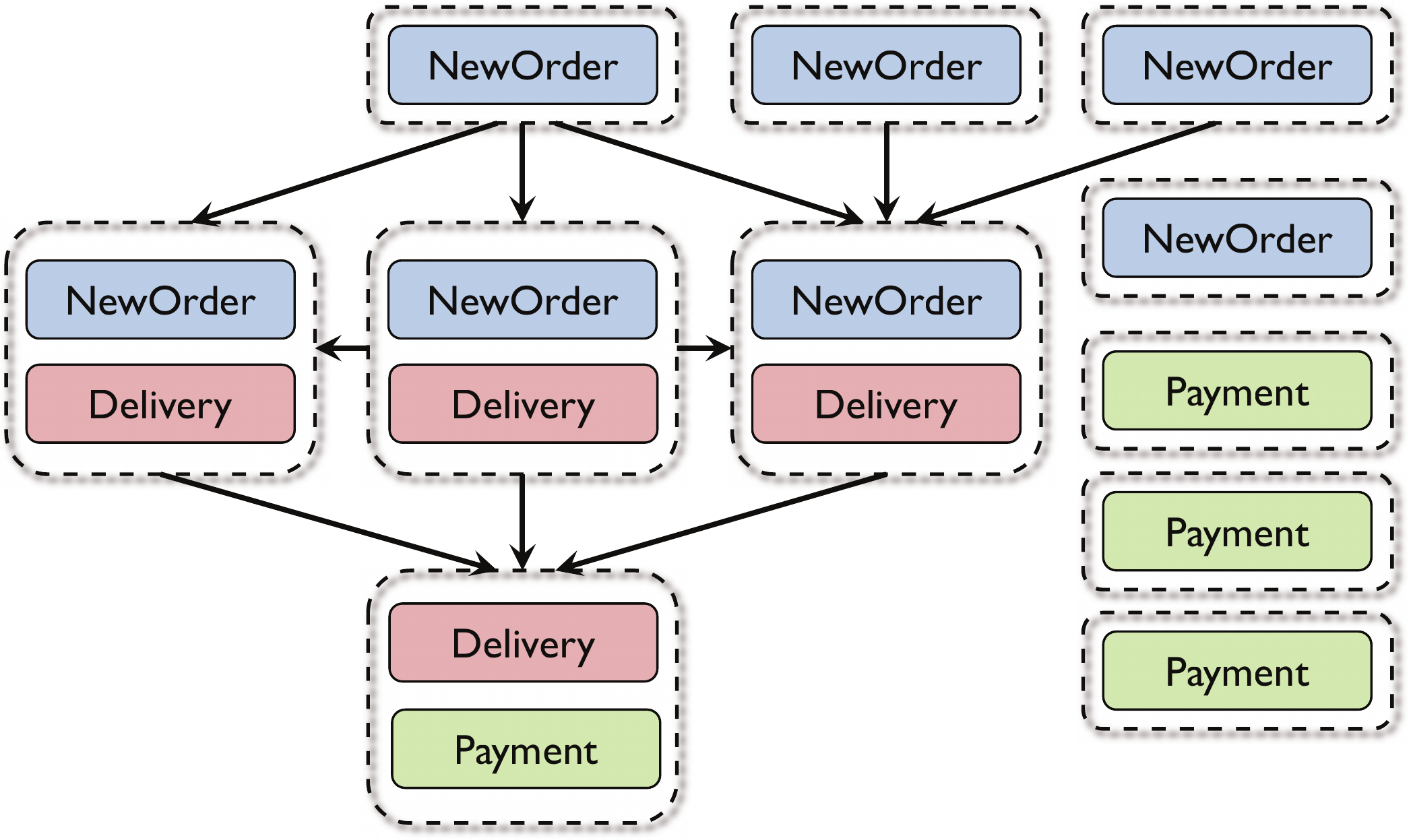}
  \caption{Global dependency graph for TPC-C. Each solid
rectangle represents a slice. Slices within the same dashed rectangle
belong to the same block.}
  \label{figure:exp:tpcc}
\end{figure}

\cref{figure:exp:tpcc} shows a simplified global dependency graph
of the TPC-C benchmark generated by \system's static analysis. 
Stored procedures in this benchmark provide a warehouse ID as an 
input parameter for each instantiated transaction. 
Note that read-only transactions are ignored as these transactions 
do not generate any logs during execution.

\section{Logging Performance}
\label{sec:logging-performance}

\begin{table}[h!]
    \centering
    \small
    \begin{tabular}{c|c|c|c|c|c|c}
    \hline
    & \multicolumn{3}{c|}{w/ checkpoint } 
    & \multicolumn{3}{c}{w/o checkpoint } \\\hhline{~------}
    & PL & LL & CL & PL & LL & CL \\\hline\hline
    \scriptsize 1 SSD (MB/s) & 352 & 347 & 250 & 274 & 252 & 34 \\
    \scriptsize 2 SSDs (MB/s) & 468 & 460 & 246 & 280 & 252 & 34 \\\hline
    \end{tabular}\vspace{5pt} 
    \caption{Overall SSD bandwidth.}
    \label{table:ssd_bandwidth}
\end{table}

\begin{table}[h!]
    \centering
    \small
    \begin{tabular}{c|c|c|c|c|c|c}
    \hline
    & \multicolumn{3}{c|}{w/ \texttt{fsync} } 
    & \multicolumn{3}{c}{w/o \texttt{fsync} } \\\hhline{~------}
    & PL & LL & CL & PL & LL & CL \\\hline\hline
    \scriptsize 1 SSD (ms) & 38 & 33 & 14 & 10 & 10 & 7 \\
    \scriptsize 2 SSDs (ms) & 25 & 24 & 11 & 10 & 10 & 7 \\\hline
    \end{tabular}\vspace{5pt} 
    \caption{Average transaction latency.}
    \label{table:fsync_latency}
\end{table}

In this section, we measure how SSD bandwidth and latency
can affect the performance of different logging schemes reported in \cref{figure:exp:logging_timeline}.

\cref{table:ssd_bandwidth} 
shows that, using one SSD, tuple-level logging (including PL and LL) 
generates approximately 350 MB/s and 260 MB/s log data
with and without checkpointing threads, respectively.
The throughput is increased to 460 MB/s when persisting data to two SSDs with 
checkpointing enabled. 
Correspondingly, we observed in \cref{figure:exp:logging_timeline}
that adding one more SSDs can greatly improve the performance of tuple-level logging
in terms of both throughput and latency.
These results altogether indicate that the throughput drops and latency
spikes observed in the experiments were due to the limitation of SSD bandwidth.
Transaction-level logging's performance is not influenced by the SSD bandwidth,
because it only generates small amounts of data. 
This is essentially a major benefit of transaction-level logging.

To analyze the effect of SSD latency,
we compare the average transaction latencies for two settings:
(1) when \texttt{fsync} is used to flush the log buffers 
(which corresponds to the latencies shown in Figure 11 in the revised version of our paper),
and (2) when \texttt{fsync} is not used at all.
\cref{table:fsync_latency} shows this comparison with checkpointing disabled.
The experiment results show that invoking \texttt{fsync} operation
can result in much higher latency for tuple-level logging (i.e., PL and LL) compared to 
transaction-level logging (i.e., CL), and
the latencies achieved by tuple-level logging can be drastically reduced 
when committing transactions without invoking \texttt{fsync} operation. 
Considering that the log size generated by tuple-level logging is $\sim$10X larger
than that of transaction-level logging,
these results altogether indicate that \texttt{fsync} is a real bottleneck 
for DBMS logging, and its overhead is exacerbated when persisting larger amounts of data.

\balance

\end{document}